\documentclass[aps,prstab,reprint,groupedaddress,amsmath,amssymb,showpacs,showkeys,floatfix]{revtex4-1}

\usepackage{graphicx}
\usepackage{dcolumn}
\usepackage{bm}

\begin{document}

\title{Dynamic aperture limitation in $e^+e^-$ colliders due to synchrotron radiation in quadrupoles
\thanks{This work has been supported by Russian Science Foundation (project  N14-50-00080).}
}

\author{ A.~Bogomyagkov }
  \email{A.V.Bogomyagkov@inp.nsk.su}
  \affiliation{Budker Institute of Nuclear Physics SB RAS, Novosibirsk 630090, Russia}
\author{ E.~Levichev }
  \affiliation{Budker Institute of Nuclear Physics SB RAS, Novosibirsk 630090, Russia}
  \affiliation{Novosibirsk State Technical University, Novosibirsk 630073, Russia}
\author{ S.~Sinyatkin }
  \affiliation{Budker Institute of Nuclear Physics SB RAS, Novosibirsk 630090, Russia}
\author{ S.~Glukhov }
  \affiliation{Budker Institute of Nuclear Physics SB RAS, Novosibirsk 630090, Russia}

\date{\today}

\begin{abstract}
In a lepton storage ring of very high energy (e.g. in the $e^+e^-$ Higgs factory) synchrotron radiation from quadrupoles constrains transverse dynamic aperture even in the absence of any magnetic nonlinearities. This was observed in tracking for LEP and the Future Circular $e^+e^-$ Collider (FCC-ee). 
Here we describe a new mechanism of instability created by modulation of the particle energy at the double betatron frequency by synchrotron radiation in the quadrupoles. Energy modulation varies transverse focusing strength at the same frequency and creates a parametric resonance of the betatron oscillations with unusual properties. It occurs at arbitrary betatron frequency (the resonant detuning is always zero) and the magnitude of the parameter modulation of the betatron oscillation (strength of the resonance driving term) depends on the oscillation amplitude. Equilibrium between the radiation damping and the resonant excitation gives the boundary of the stable motion. 
Starting from 6d equations of motion we derive and solve the relevant differential equation describing the resonance, and show good agreement between analytical results and numerical simulation.
\end{abstract}


\maketitle

\section{Introduction}
\begin{figure}[!htb]
\centering
\includegraphics*[width=.95\columnwidth,trim=20 360 10 45, clip]{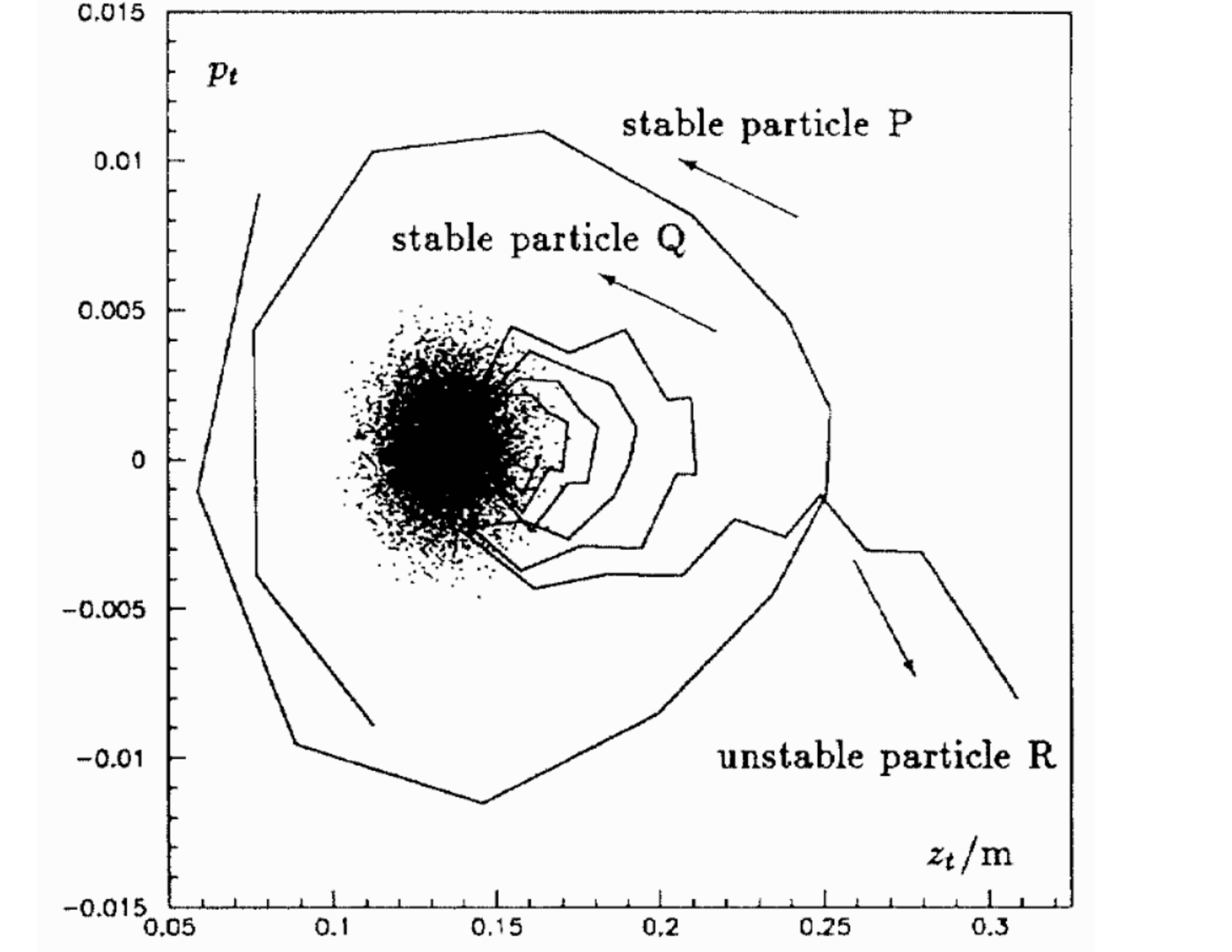}
\caption{The vertical RBSC instability in LEP at 90 GeV projected into synchrotron phase space. Three lines show the motion of three particles P, Q and R with different initial conditions. P starts with zero betatron amplitude and large longitudinal deviation. It remains stable and damps to the equilibrium synchrotron phase. Q and R start with longitudinal coordinates corresponding to the closed orbit but with vertical amplitude 5.5 mm and 6 mm respectively. Q is stable while R’s amplitude grows in few turns until it is lost. A fourth particle has been tracked with quantum emission to give the cloud of points representing the core of the beam around the closed orbit.}
\label{fig:Jowett-1}
\end{figure}

Two future electron-positron colliders FCC-ee (CERN) \cite{FCC1} and CEPC (IHEP, China)\cite{CEPC1} are now under development to carry experiments in the center-of-mass energy range from 90 GeV to 350 GeV. In these projects strong synchrotron radiation (power $\mathcal{P}\propto E^4$) is a source of effects negligible at low energy but essential at high energy, which influence beam dynamics and collider performance. One example is luminosity degradation caused by the particle radiation in the collective field of the opposite bunch (beamstrahlung \cite{Augustin:1978ah}) either due to the particle loss \cite{Telnov:2012rm} or because of the beam energy spread increase \cite{Bogomyagkov:2013uja}. Another example is about reduction of the transverse dynamic aperture due to synchrotron radiation from quadrupole magnets. John Jowett is the first who pointed out this effect in LEP collider with maximum beam energy about 100 GeV \cite{Jowett:1994yt}. Switching on the radiation from quadrupoles in the particle tracking decreased the stable betatron amplitude as compared to the radiation from bending magnets only. Jowett gave a description of this effect: ``Here I shall briefly describe a new effect which I propose to call Radiative Beta-Synchrotron Coupling (RBSC). It is a non-resonant effect. A particle with large betatron amplitude makes an extra energy loss by radiation in quadrupoles. If you imagine that its betatron amplitude does not change much over a number of synchrotron oscillations (that is not essential to the effect), you can say that its “effective stable phase angle” will change to reflect the greater energy loss. The particle will tend to oscillate about a displaced fixed point in the synchrotron phase plane. This results in a growth of the oscillation amplitude which may eventually lead the particle outside the stable region in synchrotron phase space." Jowett illustrates above assertion with synchrotron phase trajectories for two stable particles (denoted by P and Q in Figure \ref{fig:Jowett-1}) and one unstable (denoted by R) \cite{Barbarin:1994gy}. The tracking incorporates only radiation damping (quantum noise is absent) from both bending and quadrupole magnets.

In \cite{Jowett:1998au} Jowett has mentioned that the RBSC rarely occurs in isolation: ``Most often some other effect limits the dynamic aperture before the RBSC limit is reached. In the standard (LEP) lattice the horizontal dynamic aperture is limited by a rather strong shift of the vertical tune with the horizontal action variable, bringing Qy down onto the integer.”

Our interests to the subject was inspired by the FCC-ee lattice study. With the help of SAD accelerator design code \cite{SAD} K.~Oide demonstrated FCC-ee transverse dynamic aperture reduction due to radiation from quadrupoles \cite{Oide:FCC2016}, ``While the radiation loss in dipoles improves the aperture, especially at $t\bar{t}$, due to the strong damping, the radiation loss in the quadrupoles for particles with large betatron amplitudes reduces the dynamic aperture. This is due to the induced synchrotron motion through the radiation loss”. 

We crosschecked the simulation made by Oide using MAD-X PTC \cite{MADX} and the homemade software TracKing \cite{TracKing} including SR from quadrupoles and found good agreement between all three codes. Nevertheless, detailed consideration has shown different nature of the particle loss in horizontal and vertical planes. Radiation from quadrupoles at large horizontal amplitude indeed greatly shifts the synchronous phase, induces large synchrotron oscillation, excites strong synchro-betatron resonances and, finally, moves the horizontal tune toward the integer resonance (due to the nonlinear chromatic and geometrical aberrations) according to the mechanism described by Jowett and Oide. However, in the vertical plane the picture of the particle loss was quite different. The energy loss from radiation in quadrupoles for the vertical plane is substantially smaller than for the horizontal plane and does not provide large displacement of the synchronous phase and synchrotron oscillation. Instead, we found that increase of the vertical betatron oscillation amplitude modifies the vertical damping until, at some threshold, the damping changes to rising and the particle gets lost.

This new effect is a parametric resonance in oscillations with friction; radiation from quadrupoles modulates the particle energy at the double betatron frequency; therefore, quadrupole focusing strength also varies at the doubled betatron frequency creating the resonant condition. However, due to friction, resonance develops only if oscillation amplitude is larger than a certain value. The remarkable property of this resonance is that it occurs at any betatron tune (not exactly at half-integer) and hence can be labeled as ``self-inducing parametric resonance”. 

We will derive particle equations of motion in presence of the radiation from quadrupoles, consider particle loss for both transverse planes and compare results with computer simulation.

\section{Parameters values and observations from tracking}
\begin{figure*}[!htb]
\centering
\includegraphics*[width=.3\textwidth,trim=0 0 0 40, clip]{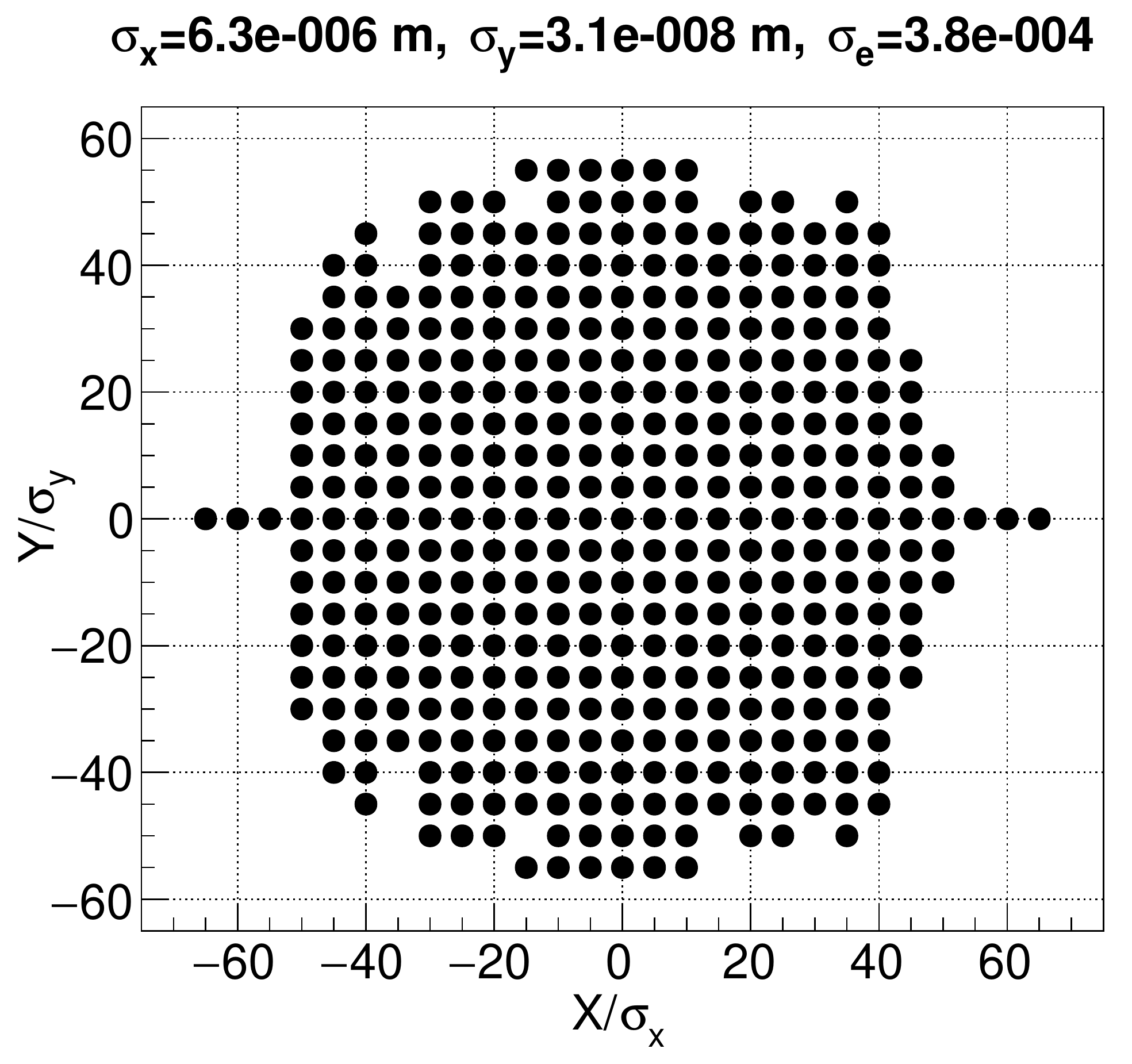}
\hfill
\includegraphics*[width=.3\textwidth,trim=0 0 0 40, clip]{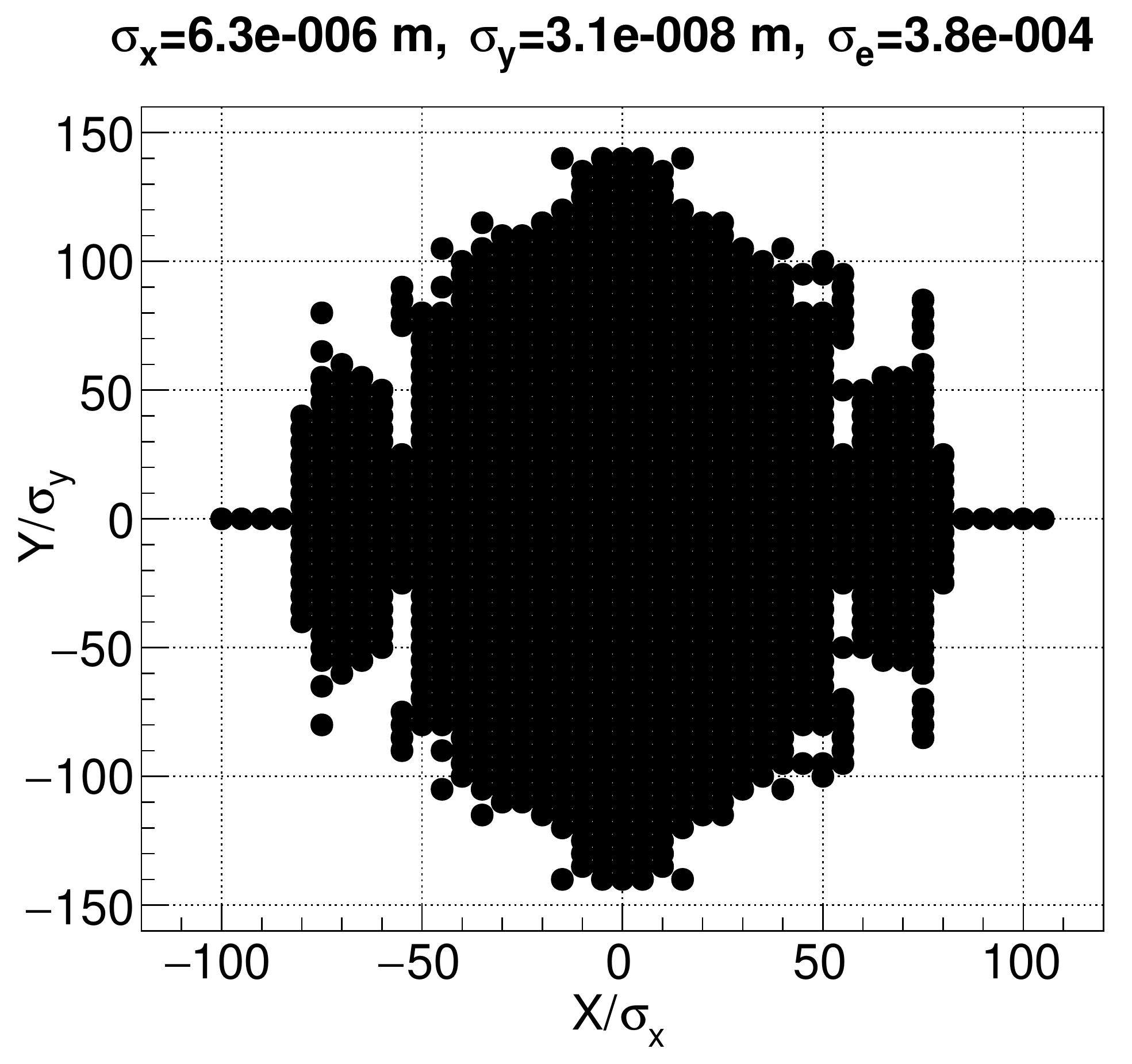}
\hfill
\includegraphics*[width=.3\textwidth,trim=155 80 155 110, clip]{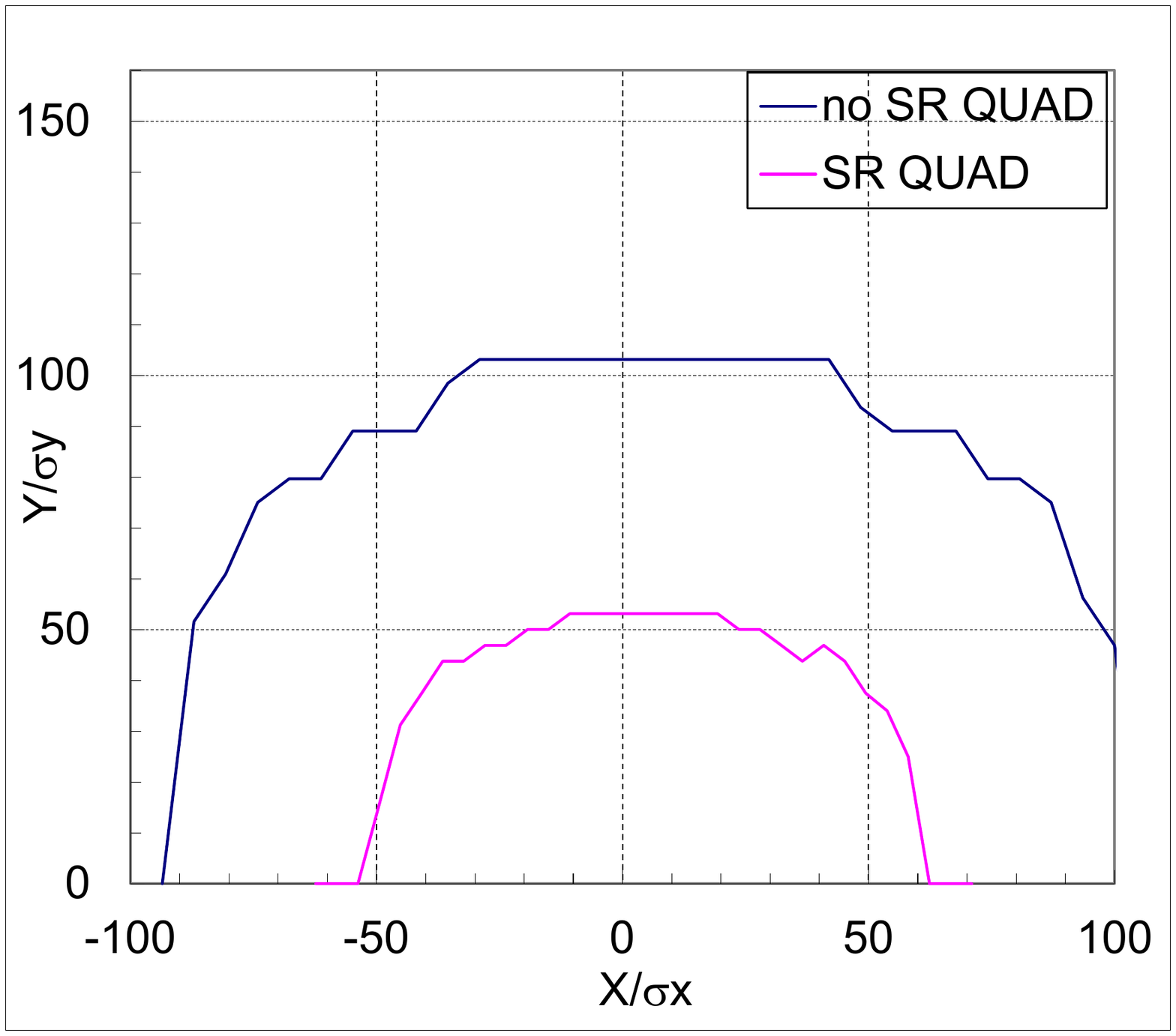}
\\
\caption{Dynamic aperture: left --- tracking by MADX PTC with synchrotron radiation from all magnetic elements, center --- tracking by MADX PTC without synchrotron radiation from all magnetic elements, right --- tracking by homemade software with synchrotron radiation from quadrupoles (blue) and without (magenta)}
\label{fig:DA-ptc-sron-0}
\end{figure*}

For the FCC-ee lattice ``FCCee\_z\_202\_nosol\_13.seq''  at 45~GeV Figure \ref{fig:DA-ptc-sron-0} shows dynamic aperture obtained by MADX PTC \cite{MADX} tracking with synchrotron radiation from all magnetic elements and without, and obtained by homemade software (TracKing \cite{TracKing}) tracking with synchrotron radiation from dipoles only and with radiation from dipoles and quadrupoles. The observation point is interaction point (IP).

Inclusion of synchrotron radiation in quadrupoles into tracking software decreases dynamic aperture
\begin{itemize}
\item in vertical direction from $R_y=142\sigma_y$ to $R_y=57\sigma_y$,
\item in horizontal direction from $R_x=109\sigma_x$ to $R_x=65\sigma_x$.
\end{itemize}

FCC-ee lattice has two IPs and Table \ref{tbl:FCC-Parameters} gives the parameters relevant to our study.
\begin{table}[!htb]
\centering
\caption{FCC-ee lattice parameters}
\begin{tabular}{|l|c|} \hline
$E_0$ [Gev]																	&$45.6$															\\ \hline
tunes: $\nu_x/\nu_y/\nu_s$											& $269.14/267.22/0.0413$								\\ \hline
damping times:																& 																		\\
$\tau_x/\tau_y/\tau_\sigma$ [turns]								& $2600/2600/1300$										\\ \hline
IP: $\beta_x/\beta_y$ [m]												& $0.15/0.001$													\\ \hline
$\varepsilon_x/\varepsilon_y$ [m]									& $2.7\times 10^{-10}/9.6 \times 10^{-13}$	\\ \hline
IP: $\sigma_x/\sigma_y$ [m]											& $6.3\times 10^{-6}/3.1\times 10^{-8}$			\\ \hline
$\sigma_\delta$															&$3.8\times 10^{-4}$										\\ \hline
\end{tabular}
\label{tbl:FCC-Parameters}
\end{table}

Table \ref{tbl:RadLoss} lists total synchrotron radiation energy loss from different type of magnets. For particles with vertical amplitude energy loss in final focus (FF) quadrupoles dominates the loss in the arc quadrupoles. For particles with horizontal amplitude energy losses in FF and in the arc quadrupoles are comparable and significantly larger than for vertical amplitudes.
\begin{table}[!htb]
\centering
\caption{Total energy loss from dipoles, final focus quadrupoles $QFF$, focusing and defocusing arc quadrupoles $QF$ and $QD$}
\begin{tabular}{|l|c|c|c|} \hline
Type				&	N			& $U(50 \sigma_x)$, MeV		& $U(50 \sigma_y)$, MeV	\\ \hline
Dipoles			& 2900		& \multicolumn{2}{|c|}{35.96}										\\ \hline
QFF				& 8			& 12										& 2									\\ \hline
QF				& 1470		& 4.1									& $3.7\times 10^{-3}$		\\ \hline
QD				& 1468		& 1.5									& $1.5\times 10^{-2}$		\\ \hline
\end{tabular}
\label{tbl:RadLoss}
\end{table}

Averaged over betatron phases radiation from quadrupoles is 
\begin{equation}
\begin{split}
U_q&=\frac{C_\gamma}{2\pi}E_0^4\oint K_1^2 (x^2+y^2) ds \\
&=E_0 \Gamma \Pi \left[\left< K_1^2 \beta_x\right>J_x+\left< K_1^2 \beta_y\right>J_y \right]\,,
\end{split}
\end{equation}
where $\Gamma=\frac{C_\gamma}{2\pi}\frac{E_0^4}{p_0c}$ is radiation related factor, $\Gamma=1.3$~m at $E_0=45.6$~GeV, $\Pi$ is circumference, angular brackets denote averaging over circumference $\left<\dots\right>=\oint \dots ds/\Pi$, and
\begin{align*}
\left< K_1^2 \beta_x\right>&=4\times 10^{-3}\,\mbox{m}^{-3}\,, \\
\left< K_1^2 \beta_y\right>&=1.4\times 10^{-1}\,\mbox{m}^{-3}\,.
\end{align*}

\begin{figure}[!ptbh]
\centering
\includegraphics*[width=0.95\columnwidth,trim=7 5 195 0, clip]{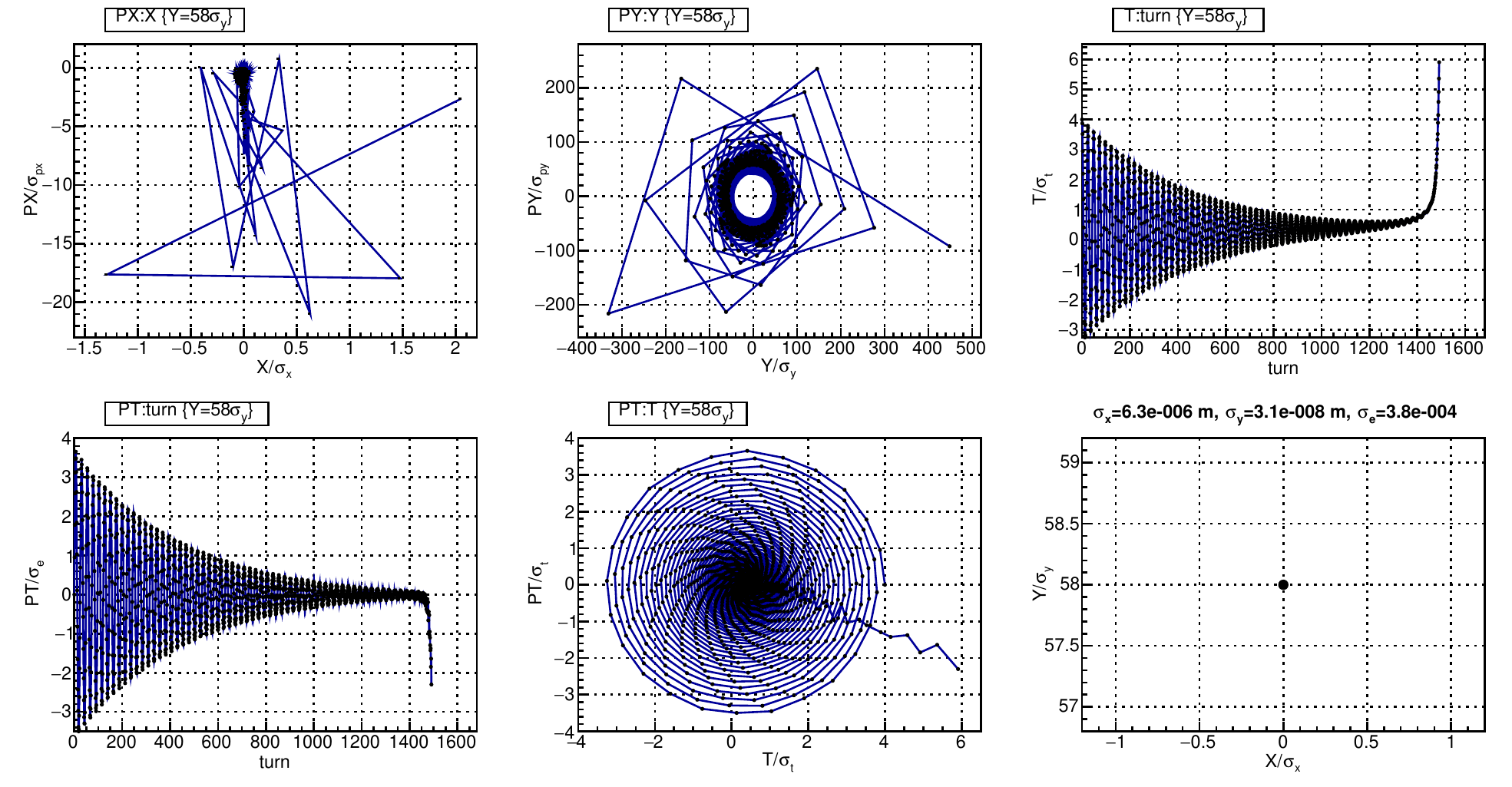}
\caption{Phase and time trajectories of the first unstable particle with initial conditions $\{x=0, y=58\sigma_y, p_x=0, p_y=0, \sigma=0, p_\sigma=0\}$.}
\label{fig:UnstablePhaseTrajectoriesY-1}
\end{figure}

For understanding the reasons of particle loss, we studied particle trajectories, obtained from tracking, in vicinity of dynamic aperture border. Figure \ref{fig:UnstablePhaseTrajectoriesY-1} shows phase and time trajectories of the first unstable (with accuracy to our step) particle with initial vertical coordinate $y=58\sigma_y$ and remaining five coordinates are zero. In the longitudinal plane $\{PT,T\}$ synchrotron oscillations excited by additional power loss from quadrupoless are damped to zero but suddenly something forces particle to walk away. Since, the longitudinal oscillations are damped they can not be the source of instability, the most probable suspect is vertical motion. In spite of initial horizontal coordinates being zero, horizontal motion is excited by nonlinear transverse coupling, however the amplitude of stable motion is not large ($<5\sigma_x$ top left plot on Fig. \ref{fig:UnstablePhaseTrajectoriesY-1}).

\begin{figure}[!ptbh]
\centering
\includegraphics*[width=0.95\columnwidth,trim=7 0 195 0, clip]{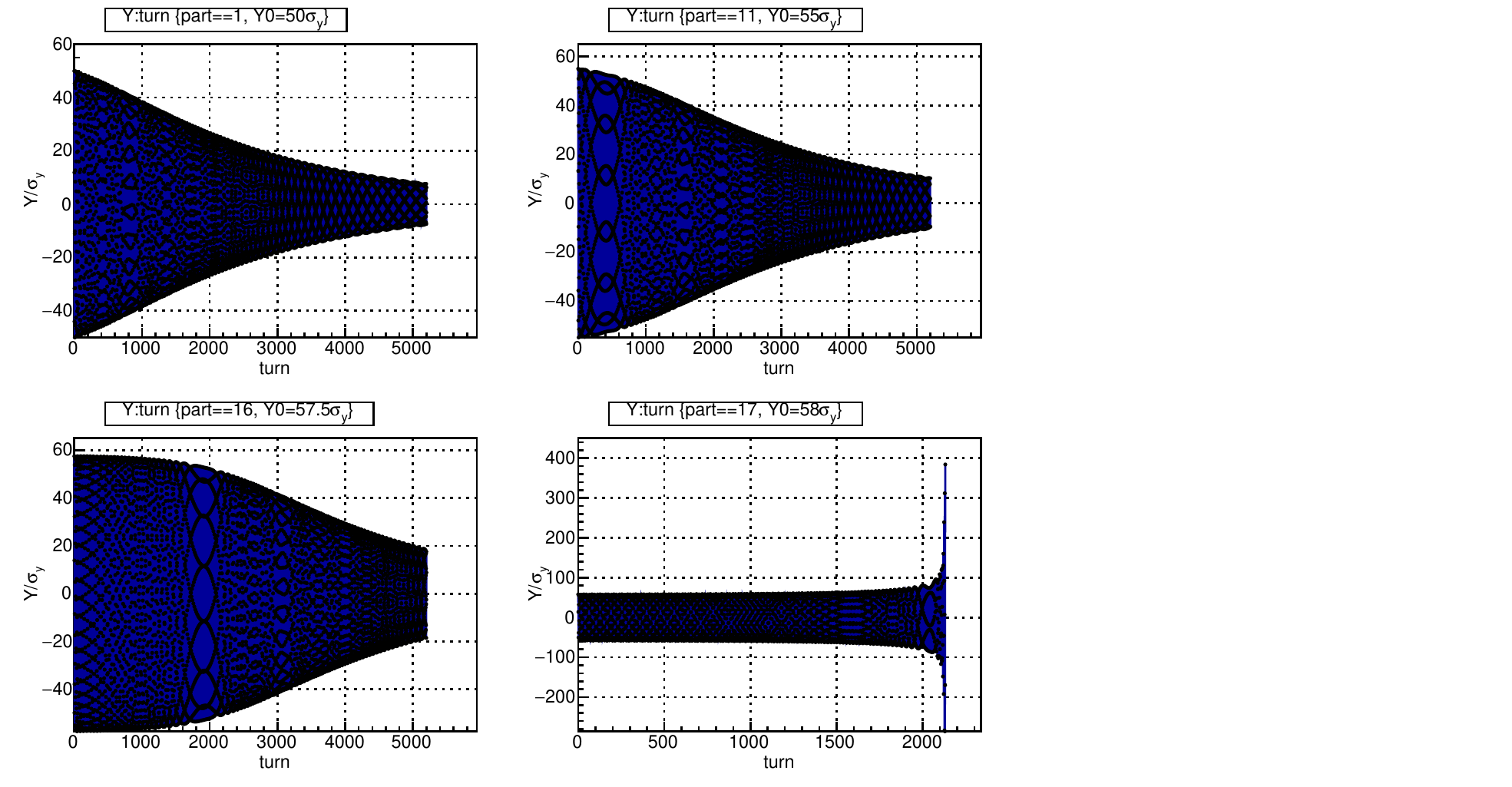}
\caption{Time evolution of vertical oscillations for particles with initial vertical coordinate $y=\{50;55;57.5;58\}\times\sigma_y$, horizontal coordinates are zero, longitudinal are adjusted for synchronous point.}
\label{fig:TrajectoriesY-1}
\end{figure}

Unexpected observations come from Figure \ref{fig:TrajectoriesY-1} showing the change of envelope evolution for particles with initial vertical coordinate around the dynamic aperture boundary $y=\{50;55;57.5;58\}\times\sigma_y$, horizontal coordinates are zero, longitudinal are chosen with respect to the new synchronous point. For the small initial amplitudes, vertical oscillations experience exponential damping, as expected, but with increase of the initial vertical amplitude and contribution of radiation power loss from quadrupoles, the envelope changes shape (left bottom plot on Fig. \ref{fig:TrajectoriesY-1}) until damping is replaced by excitation.

\begin{figure}[!ptbh]
\centering
\includegraphics*[width=0.95\columnwidth,trim=7 5 195 0, clip]{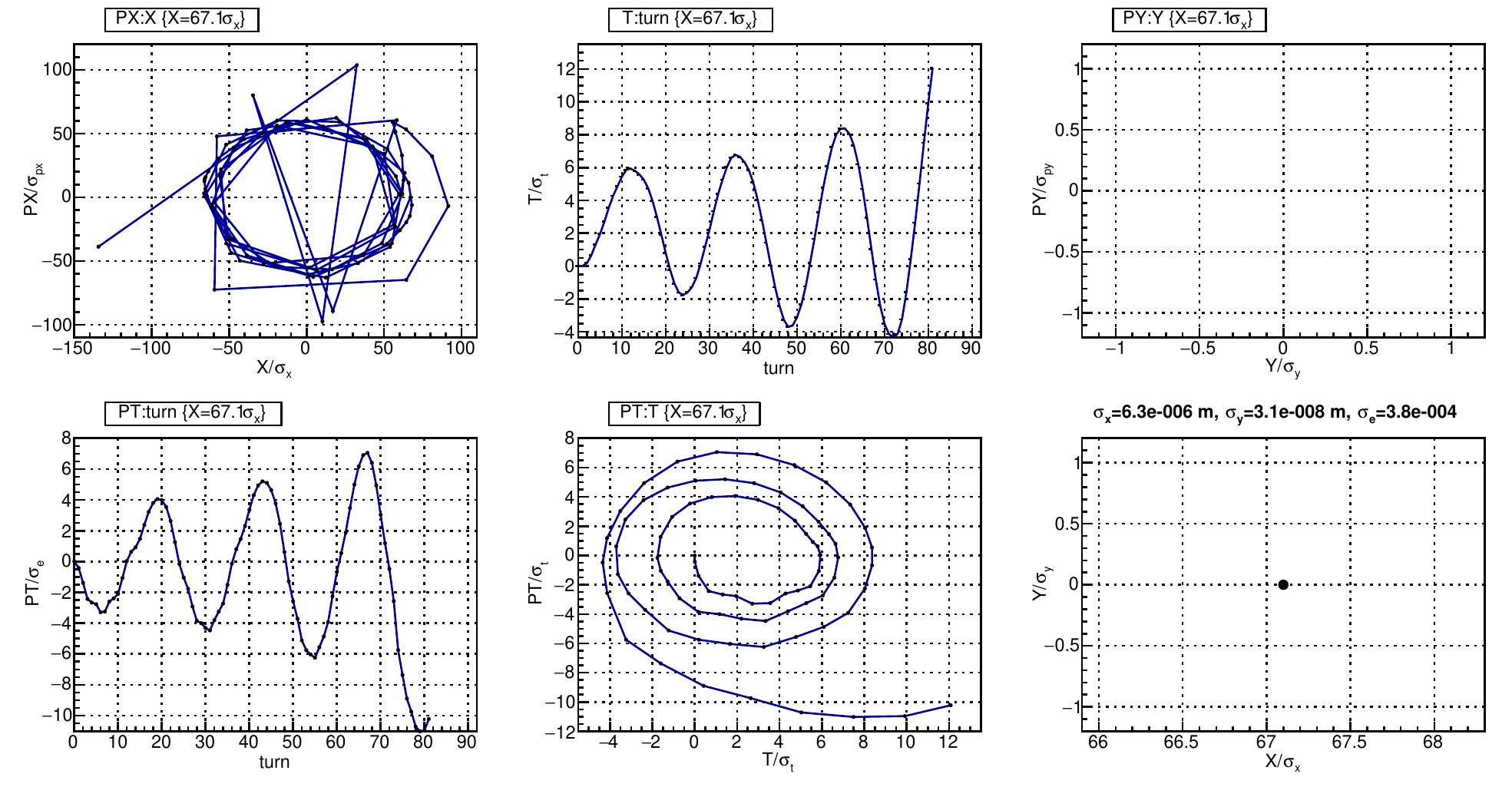}
\caption{Phase and time trajectories of the first unstable particle with initial conditions $\{x=67.1\sigma_x, y=0, p_x=0, p_y=0, \sigma=0, p_\sigma=0\}$}
\label{fig:UnstablePhaseTrajectoriesX-1}
\end{figure}

\begin{figure}[!ptbh]
\centering
\includegraphics*[width=0.95\columnwidth,trim=7 155 195 0, clip]{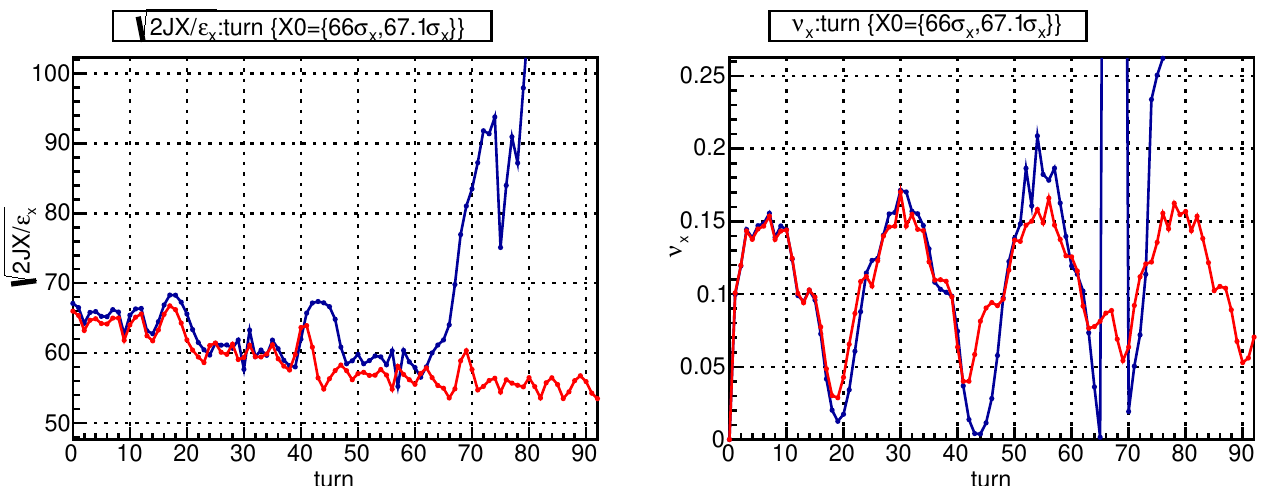}
\caption{Action and phase evolution for two particles with initial conditions: stable particle (red) with $x_0=66\sigma_x$ and unstable particle (blue) with $x_0=67.1\sigma_x$, the remaining five initial coordinates are zero. Square root of action (left). Action beating due to synchro-betatron coupling is clearly visible. Phase advance (right). Particle becomes unstable when the phase advance crosses integer value.}
\label{fig:UnstablePhaseTrajectoriesX-2}
\end{figure}

Figures \ref{fig:UnstablePhaseTrajectoriesX-1} and \ref{fig:UnstablePhaseTrajectoriesX-2} show phase and time trajectories of the first unstable particle with initial horizontal coordinate $x=67.1\sigma_x$ and remaining five zero. There is no damping and walking away in the longitudinal plane $\{PT,T\}$ as in case of vertical initial conditions Figure \ref{fig:UnstablePhaseTrajectoriesY-1}. On Figure  \ref{fig:UnstablePhaseTrajectoriesX-2} notice the right plot showing phase advance per turn with respect to turn number; the particle action starts to grow after phase advance per turn reaches an integer.

Before studing FCC-ee transverse dynamic aperture decreased by the radiation in the quadrupole magnets, we looked at the dynamic aperture caused by the lattice nonlinearities only. The transverse dynamic aperture is limited by the sextupoles for linear chromaticity correction, Maxwellian magnet fringe fields \cite{Forest:1987dr} and kinematic terms reflecting non-paraxiallity of particle motion in the first order. All chromatic sextupoles are combined in pairs with the –I optical transformation in between \cite{Oide:FCC2016}. Such arrangement cancels quadratic geometrical aberrations; therefore, the leading terms of nonlinear perturbation are cubic ones. The dynamic aperture is optimized by going through the sextupole pairs setting with a downhill simplex method scripted within SAD. It is assumed, that each sextupole pair in the arcs has individual feeding; therefore, the total optimization degrees of freedom are around 300. 

\begin{figure}[!ptbh]
\centering
\includegraphics*[width=.48\columnwidth,trim=0 0 0 0, clip]{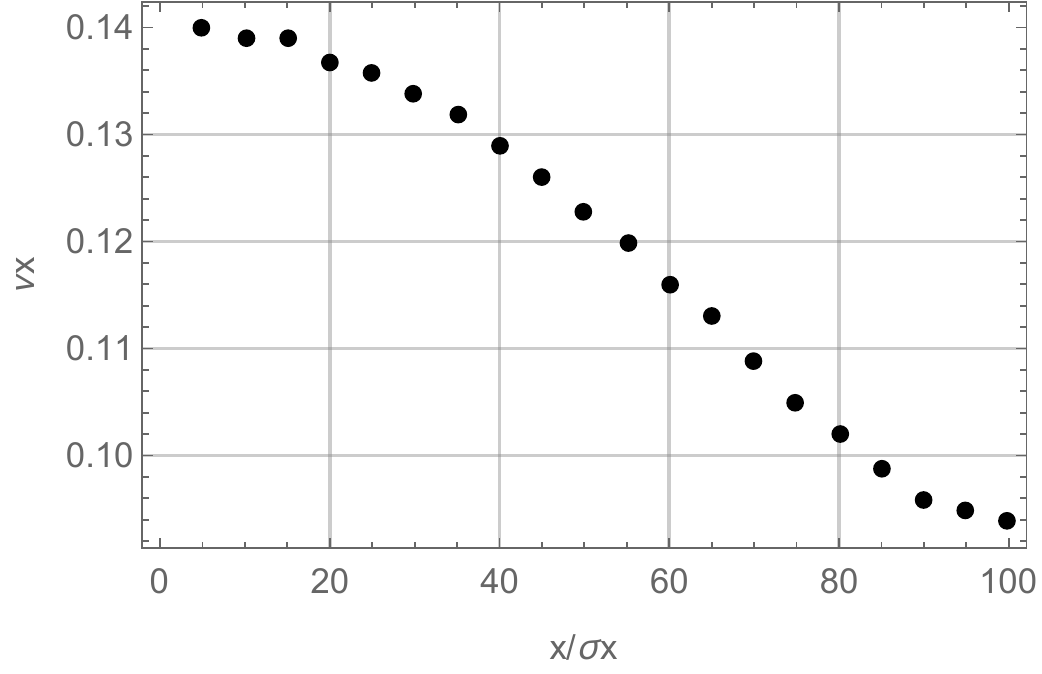}
\hfill
\includegraphics*[width=0.48\columnwidth,trim=0 0 0 0, clip]{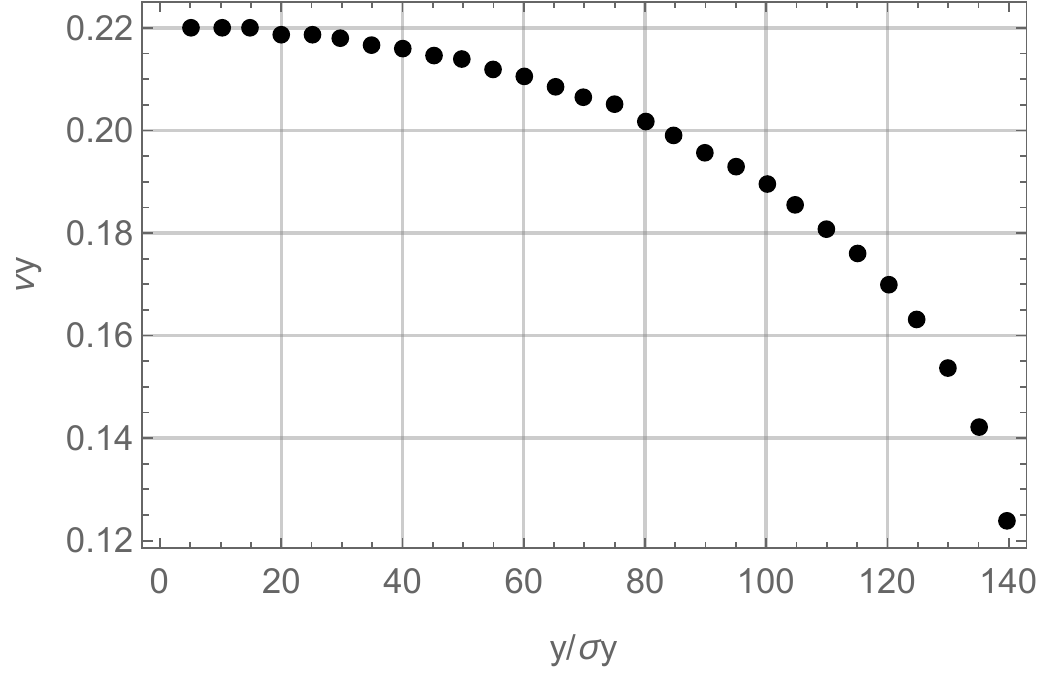}
\\
\caption{Horizontal tune dependence on initial horizontal coordinate (left) and vertical tune dependence on initial vertical coordinate (right), the remaining initial coordinates are zero.}
\label{fig:Nuxy-xy}
\end{figure}

\begin{figure}[!ptbh]
\centering
\includegraphics*[width=.48\columnwidth,trim=0 0 0 0, clip]{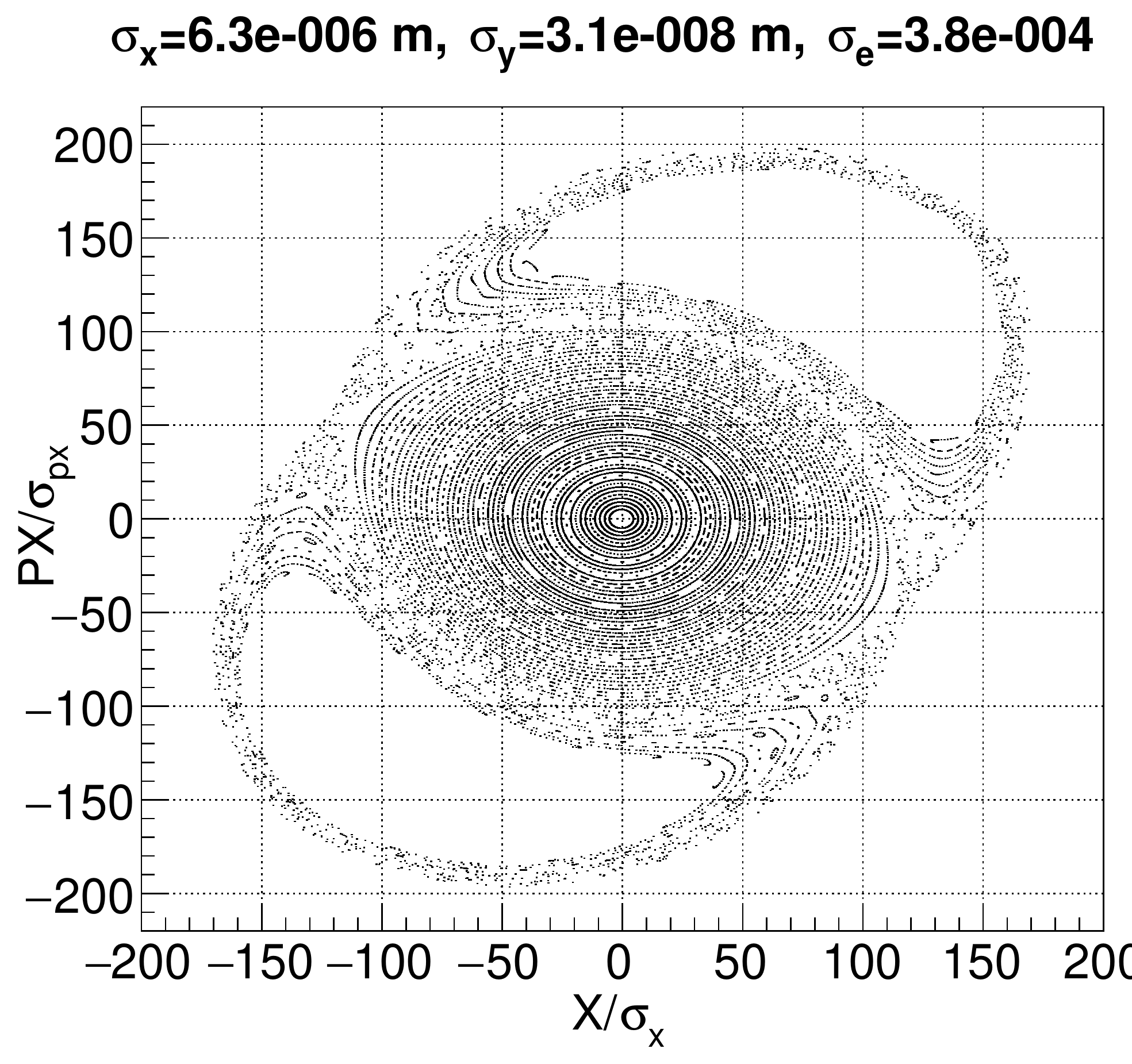}
\hfill
\includegraphics*[width=0.48\columnwidth,trim=0 0 0 0, clip]{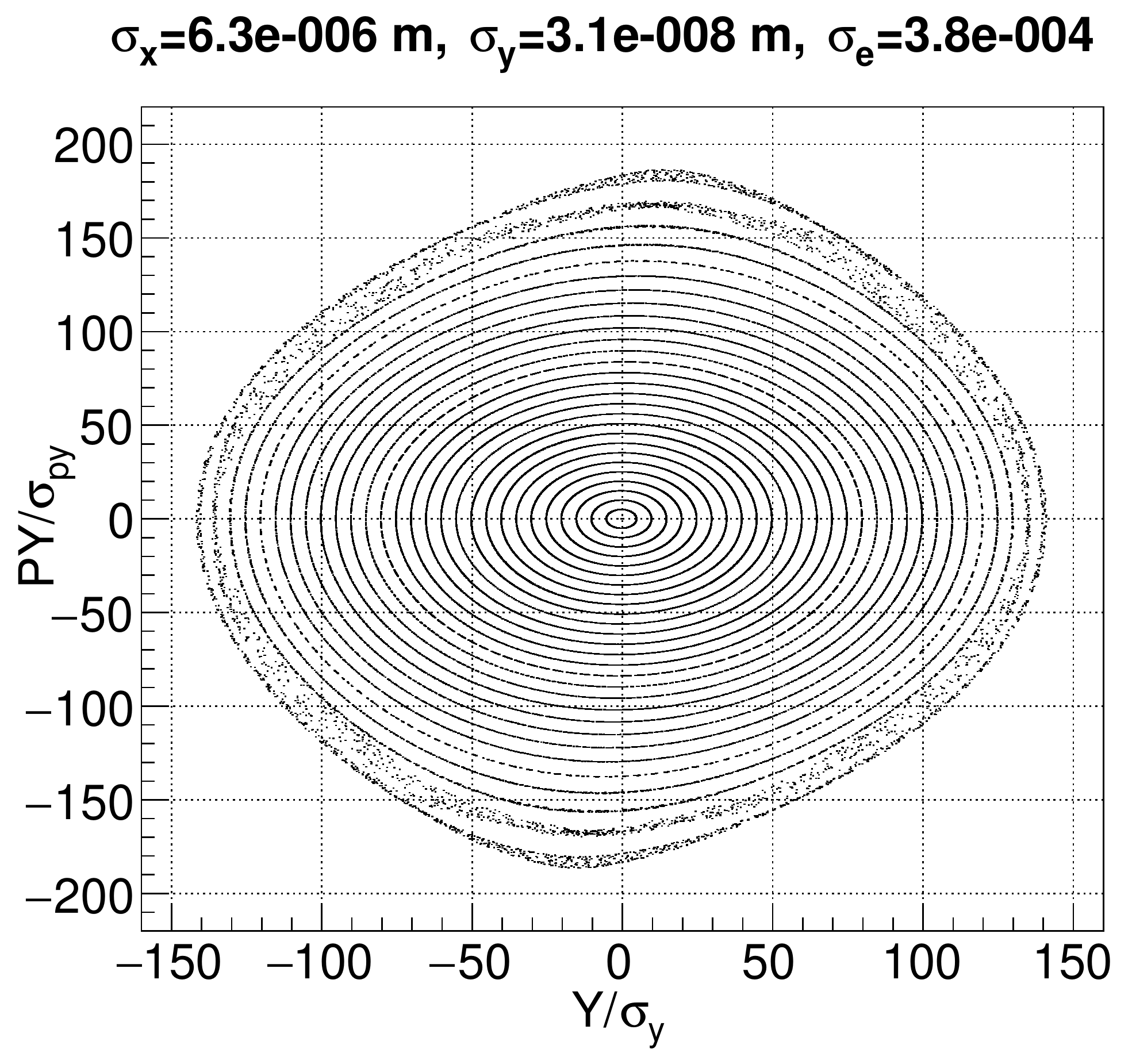}
\\
\caption{4d tracking phase trajectories: horizontal initial conditions only (left) and vertical initial conditions only (right).}
\label{fig:XY-4d}
\end{figure}

Figure \ref{fig:Nuxy-xy} shows betatron tunes as functions of initial amplitude. Both tunes move toward the nearest integer resonance $\nu_x=269$, $\nu_y=267$ with increase of initial amplitude.
However, due to the symmetry of the potential, cubic nonlinearity does not produce integer resonance. The shape of the phase trajectories on FIG. \ref{fig:XY-4d} indicates two hyperbolic fixed points in the both plots and two resonant islands in the horizontal plane, these are the signs of half-integer resonances $2\nu_x=538$, $2\nu_y=534$, which are intrinsic resonances of the potential.

\section{Equations of motion}
We start from Hamiltonian
\begin{equation}
\label{eq:Hamiltonian-0}
\begin{split}
H&(x,\sigma,y,p_x,p_\sigma,p_y;s)=1+p_\sigma+K_0 x+K_0^2\frac{x^2}{2}\\
&+K_1\frac{x^2-y^2}{2}+K_2\frac{x^3-3xy^2}{6}\\
&-(1+K_0x)\sqrt{(1+p_\sigma)^2-p_x^2-p_y^2}\\
&+\left(-\frac{eV_0}{p_0c} \right)\frac{\lambda_{RF}}{2\pi}\cos\left(\phi_s+\frac{2\pi\sigma}{\lambda_{RF}}\right)\delta(s-s_0)\,,
\end{split}
\end{equation}
where $c$ is the speed of light, $p_0$ and $E_0$ are the reference momentum and energy, $e$ is the electron charge, $B\rho=-e/p_0c$ is the rigidity, $K_0=B_y(0)/B\rho$ is the reference orbit curvature, $K_1=(dB_y/dx)/B\rho$ is the normalized quadrupole gradient, $K_2=(d^2B_y/dx^2)/B\rho$ is the normalized sextupole strength, $p_\sigma=\Delta E/p_0c$ is the longitudinal momentum, $p_{x,y}=P_{x,y}/p_0$ are the normalized transverse momenta, $V_0\,,\lambda_{RF}$ are the RF cavity voltage amplitude and wave length, $s$ is the azimuth along the orbit, $\sigma=s-ct$ is the longitudinal coordinate conjugate to the longitudinal momentum $p_\sigma$, $s_0$ is the position of point like RF cavity, $\phi_s$ is the phase of RF field.

Radiation power with assumption of negligible electron mass ($\beta=v/c=1\,, E=pc$) is
\begin{equation}
\label{eq:SR-0}
\begin{split}
\mathcal{P}&=c\frac{C_\gamma}{2\pi}e^2E^2B^2\\
&=c\frac{C_\gamma}{2\pi}E_0^4\left(1+2p_\sigma)(K_0^2+2K_0K_1x+K_1^2(x^2+y^2)\right)\\
&=c\frac{C_\gamma}{2\pi}E_0^4\left(K_0^2(1+2p_\sigma)+2K_0K_1x+K_1^2(x^2+y^2) \right)\,,
\end{split}
\end{equation}
where $B^2=(B_y+x dB_y/dx)^2+y^2(dB_y/dx)^2$ and we dropped terms with $p_\sigma^2$ and $4K_0K_1xp_\sigma$, $2K_1^2p_\sigma(x^2+y^2)$.

The next step is to expand Hamiltonian \eqref{eq:Hamiltonian-0} up to third order in all variables, neglect the term $K_0x(p_x^2+p_y^2)/2$ due to its smallness, and obtain equations of motion where radiation is included by hand with the term describing the change of momenta,
\begin{align}
\label{eq:MotionEquation-X-0}
x'&=p_x-p_xp_\sigma \\
\label{eq:MotionEquation-PX-0}
p_x'&=K_0p_\sigma-x(K_0^2+K_1)-K_2\frac{x^2-y^2}{2} \nonumber\\
&-\Gamma p_x\left[K_0^2(1+2p_\sigma)+x(2K_0K_1+K_0^3)+K_1^2(x^2+y^2) \right] \\
\label{eq:MotionEquation-Z-0}
y'&=p_y-p_yp_\sigma \\
\label{eq:MotionEquation-PZ-0}
p_y'&=yK_1+K_2xy \nonumber\\
&-\Gamma p_y\left[K_0^2(1+2p_\sigma)+x(2K_0K_1+K_0^3)+K_1^2(x^2+y^2) \right] \\
\label{eq:MotionEquation-Sigma-0}
\sigma'&=-K_0x-\frac{p_x^2}{2}-\frac{p_y^2}{2} \\
\label{eq:MotionEquation-PSigma-0}
p_\sigma'&=\left(-\frac{eV_0}{p_0c}\right)\left(\sin\phi_s+\frac{2\pi\sigma}{\lambda_{RF}}\cos\phi_s \right)\delta(s-s_0) \nonumber\\
&-\Gamma \left[K_0^2(1+2p_\sigma)+x(2K_0K_1+K_0^3)+K_1^2(x^2+y^2) \right]\!,
\end{align}
where $\Gamma=\frac{C_\gamma}{2\pi}\frac{E_0^4}{p_0c}$, and we expanded RF related $\cos(\dots)$ to first order of $\sigma$.
Note, that radiation from quadrupoles produces nonlinear terms $\Gamma K_1^2 p_{x,y} x^2$, $\Gamma K_1^2 p_{x,y} y^2$ in \eqref{eq:MotionEquation-PX-0} and \eqref{eq:MotionEquation-PZ-0} similar to the ones produced by quadrupole fringe \cite{Forest:1987dr}. However, their influence is small in our case and we drop them.

\section{Solution of longitudinal equations of motion}
\label{sect:longitudinal}
At first, we will solve longitudinal equations of motion \eqref{eq:MotionEquation-Sigma-0} and \eqref{eq:MotionEquation-PSigma-0} considering motion in the vertical plane and neglecting motion in the horizontal plane. Due to the fact that longitudinal motion is much slower than transverse (synchrotron oscillation frequency is lower than betatron), we consider vertical oscillation amplitude independent of time and solve decoupled equations. 
Splitting horizontal motion into betatron part and dispersion part $x=x_\beta+\eta p_\sigma$, $p_x=p_{x\beta}+\xi p_\sigma$, neglecting betatron motion $x_\beta=0$, $p_{x\beta}=0$ yields equations
\begin{align}
\label{eq:MotionEquation-Sigma-1}
\sigma'&=-K_0\eta p_\sigma-\xi^2\frac{p_\sigma^2}{2}-\frac{p_z^2}{2} \\
\label{eq:MotionEquation-PSigma-1}
p_\sigma'&=\left(-\frac{eV_0}{p_0c}\right)\left(\sin\phi_s+\frac{2\pi\sigma}{\lambda_{RF}}\cos\phi_s \right)\delta(s-s_0) \nonumber\\
&-\Gamma \left[K_0^2+p_\sigma(2K_0^2+2K_0K_1\eta+K_0^3\eta)\right. \nonumber\\
&\left.\qquad+K_1^2(\eta^2p_\sigma^2+y^2) \right]\!.
\end{align}

Vertical motion through nonlinear coupling excites horizontal oscillations (top left on  FIG. \ref{fig:UnstablePhaseTrajectoriesY-1}), however small ($\approx 5\sigma_x$ for $y_0=58\sigma_y$), and, according to  Table \ref{tbl:PSigma-X0-Y0} (second column, multiplying by $(5/67)^2\approx 6\cdot 10^{-3}$), excited by horizontal motion longitudinal oscillations are by order of magnitude smaller than the ones produced by vertical motion directly. Hence, we omit horizontal betatron oscillations in this section. This consideration and latter numerical oscillations will prove validity of our approximation in neglecting the nonlinear transverse coupling.

Averaging of the obtained equations over the revolution period (as usually done for synchrotron motion) introduces familiar quantities:
momentum compaction
\begin{equation}
\alpha=\left<K_0\eta\right>=\frac{1}{\Pi}\oint K_0\eta ds\,,
\end{equation}
the relative energy loss from dipoles per turn
\begin{equation}
\frac{1}{\Pi}\frac{U_0}{p_0c}=\Gamma\left<K_0^2\right>\,,
\end{equation}
wave vector of synchrotron oscillations
\begin{equation}
k_s^2=\frac{\alpha}{\Pi}\left(-\frac{eV_0}{p_0c}\right)\frac{2\pi}{\lambda_{RF}}\cos\phi_s=\left(\frac{\nu_s}{R}\right)^2\,,
\end{equation}
longitudinal damping decrement
\begin{equation}
\begin{split}
2\alpha_\sigma [m^{-1}]&=\Gamma\left<(2K_0^2+2K_0K_1\eta+K_0^3\eta)\right>\\
&=\frac{U_0}{\Pi p_0c}\left(2+\frac{\oint(2K_0K_1\eta+K_0^3\eta)ds}{\oint K_0^2 ds}\right)\\
&=\frac{U_0}{\Pi p_0c}\left(2+\frac{I_4}{I_2}\right)\,,
\end{split}
\end{equation}
where $\Pi=2\pi R$ is he ring circumference, $R$ is the average radius, angular brackets denote averaging over circumference $\left<\dots\right>=\oint \dots ds/\Pi$, $\nu_s$ is the synchrotron oscillations tune, the RF field phase is chosen according to $(-eV_0)\sin\phi_s=U_0$, $I_4$ and $I_2$ are the synchrotron integrals \cite{Helm:1973radintegrals}.

The factors $\left<\xi^2\right>$ and $\left<K_1^2\eta^2\right>$ are small, and multiplication by $p_\sigma^2$ makes them even smaller; therefore, we neglect them.

In order to deal with the terms $y^2$ and $p_y^2$, we use the principal solution of the vertical motion equation \cite{ref:Courant:1997}
\begin{equation}
\begin{aligned}
\label{eq:z-principal-0}
y&=A_yf_y+A_y^*f_y^* \\
p_y&=A_yf_y'+A_y^*f_y^{*\prime}\,,
\end{aligned} 
\end{equation}
where constant amplitude $A_y$ depends on initial conditions, $f_y$ is Floquet function with following properties
\begin{gather}
f_y=\sqrt{\beta_y}e^{i\psi_y}\,,\\
\psi_y(s)=\int_0^s\frac{d\tau}{\beta_y(\tau)}\,,\\
f_yf_y^{*\prime}-f'_yf_y^{*}=-2i\,,\\
f_y'=\frac{1}{\sqrt{\beta_y}}\left(\frac{\beta_y'}{2}+i\right)e^{i\psi_y}\,,\\
f_y'f_y^{*\prime}=\frac{1}{\beta_y}\left[\left(\frac{\beta_y'}{2}\right)^2+1\right]=\gamma_y\,,\\
f_y^{\prime 2}=\frac{1}{\beta_y}\left[\left(\frac{\beta_y'}{2}\right)^2-1+i\beta_y'\right]e^{i2\psi_y}\,,
\end{gather}
where $i$ is imaginary unit, $\beta_y$ is beta function, $\psi_y$ is betatron phase advance.
Hence, 
\begin{equation}
\begin{aligned}
y^2&=(A_yf_y+A_y^*f_y^*)^2=J_y\beta_y+A_y^2f_y^2+A_y^{*2}f_y^{*2}\,,\\
p_y^2&=(A_yf_y'+A_y^*f_y^{*\prime})^2=J_y\gamma_y+A_y^2f_y^{\prime 2}+A_y^{*2}f_y^{*\prime 2}\,,
\end{aligned}
\end{equation}
where action relates to amplitudes as $J_y=2A_yA_y^*$, Twiss parameter gamma is $\gamma_y=(1+\alpha_y^2)/\beta_y$, $\alpha_y=-\beta_y'/2$ and the subscript prime $'$ denotes $d/ds$.

In order to use Krylov-Bogolyubov averaging method we expand $p_y^2$ and $\Gamma K_1^2y^2$ into Fourier series:
\begin{align}
\Gamma K_1^2y^2&=\Gamma K_1^2\beta_yJ_y+\Gamma A_y^2 e^{i2k_ys}\sum_{n=-\infty}^{\infty}F_{y,n}e^{in\frac{s}{R}} \nonumber\\
&\quad +\Gamma A_y^{*2} e^{-i2k_ys}\sum_{n=-\infty}^{\infty}F_{y,n}^*e^{-in\frac{s}{R}}\,, \\
p_y^2&=J_y\gamma_y+A_y^2e^{i2k_ys}\sum_{n=-\infty}^{\infty}P_{y,n}e^{in\frac{s}{R}} \nonumber\\
&\quad +A_y^{*2} e^{-i2k_ys}\sum_{n=-\infty}^{\infty}P_{y,n}^*e^{-in\frac{s}{R}}\,,
\end{align}
where $k_y=2\pi\nu_y/\Pi=\nu_y/R$ is a wave vector of vertical betatron oscillations with tune $\nu_y$,
\begin{equation}
\label{eq:harmonicF}
\begin{split}
F_{y,n}&=\frac{1}{\Pi}\int_0^\Pi K_1^2(s)f_y^2(s)e^{-i2k_ys-in\frac{s}{R}}ds\\
&=\frac{1}{\Pi}\int_0^\Pi K_1^2(s)\beta_y(s)e^{i\left(2\psi_y(s)-2\nu_y\frac{s}{R}-n\frac{s}{R}\right)}ds\,,
\end{split}
\end{equation}
\begin{equation}
\label{eq:harmonicP}
\begin{split}
P_{y,n}&=\frac{1}{\Pi}\int_0^\Pi f_y^{\prime 2}(s)e^{-i2k_ys-in\frac{s}{R}}ds\\
&=\frac{1}{\Pi}\int_0^\Pi \frac{1}{\beta_y(s)}\left[\left(\frac{\beta_y'(s)}{2}\right)^2-1+i\beta_y'(s)\right] \times \\
&\quad \times e^{i\left(2\psi_y(s)-2\nu_y\frac{s}{R}-n\frac{s}{R}\right)}ds\,.
\end{split}
\end{equation}

Applying averaging method and keeping constant and slowly oscillating terms (Jowett kept constant, but omitted oscillating terms in \cite{Jowett:1986pc}) yields equations of motion
\begin{align}
\sigma'&= -\alpha p_\sigma -J_y\frac{\left<\gamma_y\right>}{2} \nonumber \\
\label{eq:MotionEquation-Sigma-2}
&\qquad-\frac{A_y^2}{2}P_{y,n}e^{i\frac{s}{R}(2\nu_y+n)} \\
&\qquad-\frac{A_y^{*2}}{2}P_{y,n}^*e^{-i\frac{s}{R}(2\nu_y+n)} \nonumber\,,\\
p_\sigma'&=\frac{k_s^2}{\alpha}\sigma-2\alpha_\sigma p_\sigma-\Gamma \left<K_1^2\beta_y\right>J_y \nonumber\\
\label{eq:MotionEquation-PSigma-2}
&\qquad-\Gamma A_y^2 F_{y,n}e^{i\frac{s}{R}(2\nu_y+n)} \\
&\qquad-\Gamma A_y^{*2}F_{y,n}^*e^{-i\frac{s}{R}(2\nu_y+n)} \nonumber\,,
\end{align}
where $n=-[2\nu_y]$ is the negative integer part of the double betatron tune and is the only slow oscillating harmonic.

\subsection{Synchronous phase}
Equating the right parts of the equations \eqref{eq:MotionEquation-Sigma-2} and \eqref{eq:MotionEquation-PSigma-2} to zero and eliminating the oscillating terms results in synchronous longitudinal point
\begin{align}
\label{eq:SynchroSigmaY}
\sigma&=-\frac{\alpha_\sigma}{k_s^2}\left<\gamma_y\right>J_y+\frac{\alpha}{k_s^2}\Gamma\left<K_1^2\beta_y\right>J_y \\
\label{eq:SynchroPSigmaY}
p_\sigma&=-\frac{1}{2\alpha}\left<\gamma_y\right>J_y\,,
\end{align}
where the term with $\Gamma$ corresponds to additional energy loss from radiation in quadrupoles, the other terms come from lengthening of particle trajectory. Jowett obtained similar equations in \cite{Jowett:1994yt} and \cite{Jowett:1986yx}.

Particle with not adjusted initial conditions will develop synchrotron oscillations with respect to the new synchronous point. Using the longitudinal invariant
\begin{equation}
\sigma^2+\frac{\alpha^2}{k_s^2}p_\sigma^2=const
\end{equation}
yields maximum energy deviation
\begin{equation}
\label{eq:P-Sigma-max-Z}
p_{\sigma, max}=J_y\sqrt{\left(-\frac{\alpha_\sigma\left<\gamma_y\right>}{k_s\alpha}+\frac{\Gamma\left<K_1^2\beta_y\right>}{k_s}\right)^2+\frac{\left<\gamma_y\right>^2}{4\alpha^2}}
\end{equation}

\subsection{Solution without oscillating terms}
Solution of equations \eqref{eq:MotionEquation-Sigma-2} and \eqref{eq:MotionEquation-PSigma-2} without oscillating terms is known and consists of the constant term describing the shift of synchronous energy, and two terms describing damping synchrotron oscillations (only for $p_\sigma$)
\begin{equation}
\label{eq:Solution-PSigma-1}
\begin{split}
p_\sigma&=-\frac{\left<\gamma_y\right>}{2\alpha}J_y
+B_1e^{-\alpha_\sigma s}\cos\left(s\sqrt{k_s^2-\alpha_\sigma^2}\right) \\
&\qquad+B_2e^{-\alpha_\sigma s}\sin\left(s\sqrt{k_s^2-\alpha_\sigma^2}\right)\,.
\end{split}
\end{equation}

\subsection{Particular solution}
Introducing $\mbox{\ae}_y=(2\nu_y+n)/R$ and transforming the system of first order differential equations \eqref{eq:MotionEquation-Sigma-2} and \eqref{eq:MotionEquation-PSigma-2} into the the second order equation gives
\begin{equation}
\label{eq:MotionEquation-PSigma-3}
\begin{split}
p_\sigma''&+k_s^2p_\sigma+2\alpha_\sigma p_\sigma'=  \\
& -A_y^2\left(\frac{k_s^2}{2\alpha}P_{y,n}+i\Gamma\mbox{\ae}_y F_{y,n}\right)e^{i\mbox{\ae}_y s} \\
& -A_y^{*2}\left(\frac{k_s^2}{2\alpha}P_{y,n}^*-i\Gamma\mbox{\ae}_y F_{y,n}^*\right)e^{-i\mbox{\ae}_y s}\,.
\end{split}
\end{equation}
Particular solution of \eqref{eq:MotionEquation-PSigma-3} is
\begin{equation}
\label{eq:Solution-PSigma-2}
\begin{split}
p_\sigma&=
-\frac{A_y^2\left(\frac{k_s^2}{2\alpha}P_{y,n}+i\Gamma\mbox{\ae}_y F_{y,n}\right)}{k_s^2-\mbox{\ae}_y^2+i2\mbox{\ae}_y\alpha_\sigma}e^{i\mbox{\ae}_y s}\\
&\quad-\frac{A_y^{*2}\left(\frac{k_s^2}{2\alpha}P_{y,n}^*-i\Gamma\mbox{\ae}_y F_{y,n}^*\right)}{k_s^2-\mbox{\ae}_y^2-i2\mbox{\ae}_y\alpha_\sigma}e^{-i\mbox{\ae}_y s}\,.
\end{split}
\end{equation}
Since 
\begin{gather}
\label{eq:inequality-nu}
\mbox{\ae}_y\gg k_s\gg\alpha_\sigma\,, \\
\label{eq:inequality-harmonics}
\Gamma\mbox{\ae}_y\left|F_{y,n}\right|\gg \frac{k_s^2}{2\alpha}\left|P_{y,n}\right| 
\end{gather}
we can rewrite solution as
\begin{equation}
\label{eq:Solution-PSigma-3}
p_\sigma\approx
i A_y^2\frac{\Gamma F_{y,n}}{\mbox{\ae}_y}e^{i\mbox{\ae}_y s}
-i A_y^{*2}\frac{\Gamma F_{y,n}^*}{\mbox{\ae}_y}e^{-i\mbox{\ae}_y s}\,.
\end{equation}
Apparently, solutions \eqref{eq:Solution-PSigma-2} and \eqref{eq:Solution-PSigma-3} should not depend on the initial betatron phase $\varphi_y$, because in the averaging over the revolution period we lose all the information regarding particle initial transverse phase. Therefore, we replace complex betatron amplitude  $A_y=\left|A_y\right|\exp(i \varphi_y)$ with its absolute value $\left|A_y\right|$. Putting it in the form comfortable for the future use we have
\begin{equation}
\label{eq:Solution-PSigma-4}
\begin{split}
p_\sigma&=
c_n \left|A_y\right|^2e^{i\mbox{\ae}_y s}+c_n^*\left|A_y\right|^2e^{-i\mbox{\ae}_y s}\\
&=\left|c_n\right|J_y\cos\left(\mbox{\ae}_y s+\chi_0\right)\,,
\end{split}
\end{equation}
where 
\begin{equation}
\label{eq:harmonicC}
c_n=-\frac{\left(\frac{k_s^2}{2\alpha}P_{y,n}+i\Gamma\mbox{\ae}_y F_{y,n}\right)}{k_s^2-\mbox{\ae}_y^2+i2\mbox{\ae}_y\alpha_\sigma}
\approx i \frac{\Gamma F_{y,n}}{\mbox{\ae}_y}
\end{equation}
and $\chi_0=\arg(c_n)$.

\section{Solution of vertical equations of motion}
With the same assumptions as in the previous paragraph equations \eqref{eq:MotionEquation-Z-0} and \eqref{eq:MotionEquation-PZ-0} are
\begin{align}
\label{eq:MotionEquation-Z-1}
y'&=p_y-p_yp_\sigma \,,\\
\label{eq:MotionEquation-PZ-1}
p_y'&=K_1 y+K_2\eta p_\sigma y -\Gamma p_y\left[K_0^2 +p_\sigma D+K_1^2 y^2 \right] \!,
\end{align}
where $D=2K_0^2+2K_0K_1\eta+K_0^3\eta$ and for machines with separate functions magnets is negligible, we neglected the small term $\Gamma p_y K_1^2 \eta^2p_\sigma^2$. We may apply Krylov-Bogolyubov averaging method directly to equations \eqref{eq:MotionEquation-Z-1}, \eqref{eq:MotionEquation-PZ-1}, but it is more illustrative to apply it to $y^{\prime\prime}$ equation. During derivation of $y^{\prime\prime}$ equation we neglect the terms containing $p_\sigma^{\prime}$, because it either oscillates with synchrotron tune or with double fractional part of betatron frequency, and after derivation will receive a small factor. The desired equation is
\begin{equation}
\label{eq:zdoubleprime-1}
y''-\left(K_1-(K_1-K_2\eta)p_\sigma\right)y+\Gamma\left(K_0^2+K_1^2y^2\right)y'=0\,.
\end{equation}
This is an equation of parametric oscillator with friction; the second term depends on $p_\sigma$ which contains terms oscillating at fractional double betatron frequency \eqref{eq:Solution-PSigma-4}. It is also a Van der Pol oscillator (nonlinear friction, the the third term). Jowett obtained Van der Pol equation for nonlinear wiggler (combined quadrupole and sextupole) in \cite{Jowett:1986yx}. We did not find large influence of nonlinear friction (Van der Pol oscillator) and, therefore, omitted it.

Substituting expression for $p_\sigma$, we neglect the constant shift and damped synchrotron oscillations \eqref{eq:Solution-PSigma-1}, and keep only particular solution \eqref{eq:Solution-PSigma-4} oscillating on fractional part of double betatron frequency, i.e. we consider only parametric resonance. Substituting principal solution for $y$ \eqref{eq:z-principal-0}, averaging and keeping only slowly oscillating terms yields equation for amplitude evolution
\begin{equation}
\label{eq:Az-1}
\begin{split}
(-2i)A_y'&=A_y\left<\Gamma K_0^2(-\alpha_y+i)\right> \\
&+ \left|A_y\right|^2A_y^*\left|c_n\right|\left<(K_1-K_2\eta)\beta_y e^{i(-2\psi_y+\mbox{\ae}_y s+\chi_0)}\right>\\
&-3A_y^2A_y^*\left<\Gamma K_1^2\beta_y\alpha_y\right>+iA_y^2A_y^*\left<\Gamma K_1^2\beta_y\right>\,.
\end{split}
\end{equation}
The terms $\left<\Gamma K_1^2\beta_y\alpha_y\right>$ and $\left<\Gamma K_1^2\beta_y\right>$ are small and we neglect them, obtaining
\begin{equation}
\label{eq:Az-2}
\begin{split}
A_y'&=-\frac{1}{2}\left<\Gamma K_0^2(1+i\alpha_y)\right>A_y \\
 &\quad+\frac{i}{2}\left|c_n\right|\left<(K_1-K_2\eta)\beta_y e^{i(-2\psi_y+\mbox{\ae}_y s+\chi_0)}\right>\left|A_y\right|^2A_y^* \\
 &=-B_1 A_y+i B_2 \left|A_y\right|^2A_y^*\,.
\end{split}
\end{equation}
The real part of the obtained equation describes evolution of the $\left|A_y\right|$ (e.g. damping), the imaginary part describes the change of the betatron tune. In order to solve equation \eqref{eq:Az-2} we introduced coefficients
\begin{align}
\label{eq:B1}
B_1&=\frac{1}{2}\left<\Gamma K_0^2(1+i\alpha_y)\right> \\
\label{eq:B2}
B_2&=\frac{1}{2}c_n\left<(K_1-K_2\eta)\beta_y e^{i(-2\psi_y+\mbox{\ae}_y s)}\right>\,,
\end{align}
where expression in angular brackets of $B_2$ is local chromaticity, which does not vanish when global chromaticity is compensated.

Distinguishing modulus and argument of amplitude $A_y=a_y e^{i\varphi_y}$, $B_1=\left|B_1\right|e^{i\varphi_1}$, $B_2=\left|B_2\right|e^{i\varphi_2}$ and substituting in \eqref{eq:Az-2} results in two equations
\begin{align}
\label{eq:Az-4}
a_y'&=-a_y\left|B_1\right|\cos(\varphi_1)-a_y^3\left|B_2\right|\sin(-2\varphi_y+\varphi_2)\,, \\
\label{eq:Phiz-4}
\varphi_y'&=-\left|B_1\right|\sin(\varphi_1)+a_y^2\left|B_2\right|\cos(-2\varphi_y+\varphi_2)\,,
\end{align}
where $\left|B_1\right|\sin(\varphi_1)=Im(B_1)=\frac{1}{2}\left<\Gamma K_0^2 \alpha_y\right>\approx 0$ is small and describes the change of vertical betatron tune because of damping; this is equivalent to $\varphi_1=0$. The second term in \eqref{eq:Phiz-4} describes tune dependence on amplitude. Equations \eqref{eq:Az-4} and \eqref{eq:Phiz-4} have complex topology in $\{a_y,\varphi_y\}$ space (see Appendix), which has two stable points providing $\varphi_y'=0$
\begin{equation}
\varphi_y=\frac{\varphi_2}{2}\pm \frac{\pi}{4} +\pi n\,,
\end{equation}
where $n$ is integer. At these points the modulus of amplitude is
\begin{equation}
a_y(s)=\frac{a_{y,0}e^{-\left|B_1\right|s}}{\sqrt{1\pm a_{y,0}^2\frac{\left|B_2\right|}{\left|B_1\right|}\left(1-e^{-\left|B_1\right|s}\right)}}\,,
\end{equation}
and using $J_y=2A_yA_y^*=2a_y^2$ gives action
\begin{equation}
\label{eq:Solution-Jz-1}
J_y(s)=\frac{J_{y,0} e^{-2\left|B_1\right| s}}{1\pm J_{y,0}\frac{\left|B_2\right|}{2\left|B_1\right|}(1-e^{-2\left|B_1\right| s})}\,.
\end{equation}
The plus sign describes always damping amplitudes (stable), the minus sign, depending on initial action, describes either damping solutions (stable) or rising (unstable). This boundary action defines the border of dynamic aperture and is
\begin{equation}
\label{eq:Jzlimit}
J_{y,lim}=\frac{2\left|B_1\right|}{\pm \left|B_2\right|}\,.
\end{equation}
Existence of initial amplitudes with stable motion at parametric resonance is due to the friction (radiation damping).

\section{Longitudinal and horizontal motion}
Equations of coupled horizontal and longitudinal motion \eqref{eq:MotionEquation-X-0}, \eqref{eq:MotionEquation-PX-0},\eqref{eq:MotionEquation-Sigma-0}, \eqref{eq:MotionEquation-PSigma-0} with $y=0$ and $p_y=0$  are similar to vertical and longitudinal \eqref{eq:MotionEquation-Z-0} \eqref{eq:MotionEquation-PZ-0} with $x_\beta=0$ $p_{x\beta}=0$. The unique for horizontal motion terms $K_0p_\sigma$ in \eqref{eq:MotionEquation-PX-0} responsible for dispersion and $-K_0x_\beta$ in \eqref{eq:MotionEquation-Sigma-0} will produce a synchro-betatron resonance at $\nu_x\pm\nu_s=integer$. This resonance plays an important role, but out of scope of our work. Table \ref{tbl:PSigma-X0-Y0} shows that the shift of synchronous point and amplitude of synchrotron oscillations are significantly larger for horizontal oscillations (second column) than for vertical (third column) at the boundary of dynamic aperture, if initial longitudinal coordinates are not adjusted to the new synchronous point.
\begin{table}[!hbt]
\centering
\caption{Synchronous point and amplitude of synchrotron oscillations for different transverse initial conditions}
\begin{tabular}{|l|c|c|} \hline
$\{X_0,Y_0\}$								& $\{67\sigma_x,0\}$	& $\{0,58\sigma_y\}$	\\ \hline
$p_{\sigma, max}/\sigma_\delta$	& $4$							& $0.26$ 						\\ \hline
$p_{\sigma, syn}/\sigma_\delta$	& $-2.5$						& $-0.026$					\\ \hline
$\sigma_{syn}/\sigma_\delta$		& $3.05$						& $0.29$						\\ \hline
\end{tabular}
\label{tbl:PSigma-X0-Y0}
\end{table}
Observation of phase advance per turn (right) on Figure~\ref{fig:UnstablePhaseTrajectoriesX-2} suggests that particle is lost when phase advance reaches an integer (turn 65) and it happens when $p_\sigma=7\sigma_\delta$. Using the detuning coefficient and its chromaticity with given initial conditions we calculated the shift of the tune from each term Table~\ref{tbl:dnu}. The sum of last three lines is exactly zero, which means that the tune is equal integer.
\begin{table}[!hbt]
\centering
\caption{Tune shift contribution from detuning and detuning chromaticity}
\begin{tabular}{|l|c|} \hline
$\displaystyle\frac{\partial \nu_x}{\partial J_x}$ 							& $-5\times 10^4$				\\ \hline
$\displaystyle\frac{\partial^2 \nu_x}{\partial J_x\partial \delta}$	& $-6.8\times 10^7$				\\ \hline
$J_x$																							& $\displaystyle 67^2\varepsilon_x\big/2$ \\ \hline
$p_\sigma$																					& $7\sigma_\delta$ \\ \hline
$\Delta\nu_x=\displaystyle\frac{\partial \nu_x}{\partial J_x} J_x$ 	& $-0.03$  \\ \hline
$\Delta\nu_x=\displaystyle\frac{\partial^2 \nu_x}{\partial J_x\partial \delta}J_x p_\sigma$ & $-0.11$  \\ \hline
$\nu_x(J_x=0,\, p_\sigma=0)$ 														& 0.14 \\ \hline
\end{tabular}
\label{tbl:dnu}
\end{table}

\section{Comparison with tracking and numerical estimations}
\subsection{Vertical motion}
For given vertical tune harmonic number is $n=-534$, $\mbox{\ae}_y=2.8\times 10^{-5}$~m$^{-1}$, $k_s=2.6\times 10^{-6}$~m$^{-1}$. The harmonics \eqref{eq:harmonicF}, \eqref{eq:harmonicP} and  \eqref{eq:harmonicC} are
\begin{align*}
F_{y,n}&=(-0.14, 3\times 10^{-5})\,\mbox{m}^{-3}	& \left|F_{y,n}\right|&=0.14\,\mbox{m}^{-3} \\
P_{y,n}&=(-0.13,0.0006)\,\mbox{m}^{-1}					& \left|P_{y,n}\right|&=0.13\,\mbox{m}^{-1} \\
c_n&=(-42.11,-6474.19)\,\mbox{m}^{-1}					& \left|c_n\right|&=6474.33\,\mbox{m}^{-1}\,,
\end{align*}
where expression in brackets $(,)$ designates real and imaginary part of the value respectfully.
The numbers prove the inequality \eqref{eq:inequality-harmonics}
\begin{gather*}
\Gamma \mbox{\ae}_y \left|F_{y,n}\right|=5.13\times 10^{-6} \\
\frac{k_s^2}{2 \alpha} \left|P_{y,n}\right|=3.22\times 10^{-8}\,.
\end{gather*}
Coefficients \eqref{eq:B1} and \eqref{eq:B2} are 
\begin{align*}
B_1&=(4.03\times 10^{-9},-2.76\times 10^{-10})\,\mbox{m}^{-1}	\\
\left|B_1\right|&=4.04\times 10^{-9}\,\mbox{m}^{-1}		\\
B_2&=(10.35,6.43)\,\mbox{m}^{-2}		\\
\left|B_2\right|&=12.18\,\mbox{m}^{-2}\,.
\end{align*}
The border of dynamic aperture \eqref{eq:Jzlimit} is
\begin{equation}
\label{eq:Ylimit-number}
R_y=\sqrt{2J_{y,lim}\beta_y}=37.2\sigma_y\,,
\end{equation}
which needs to be compared with the tracking result $R_y=57\sigma_y$. Scrutiny of tracking results showed that transverse nonlinear coupling decreases effective amplitude of vertical motion; therefore, the amplitude of longitudinal harmonic producing parametric resonance is about two times smaller than our predictions. Consideration of this correction increases dynamic aperture $R_y\approx37.2\times\sqrt{2}\sigma_y=52.6\sigma_y$, which corresponds well to tracking results.

\begin{figure}[!htb]
\centering
\includegraphics*[width=.95\columnwidth,trim=0 0 0 0, clip]{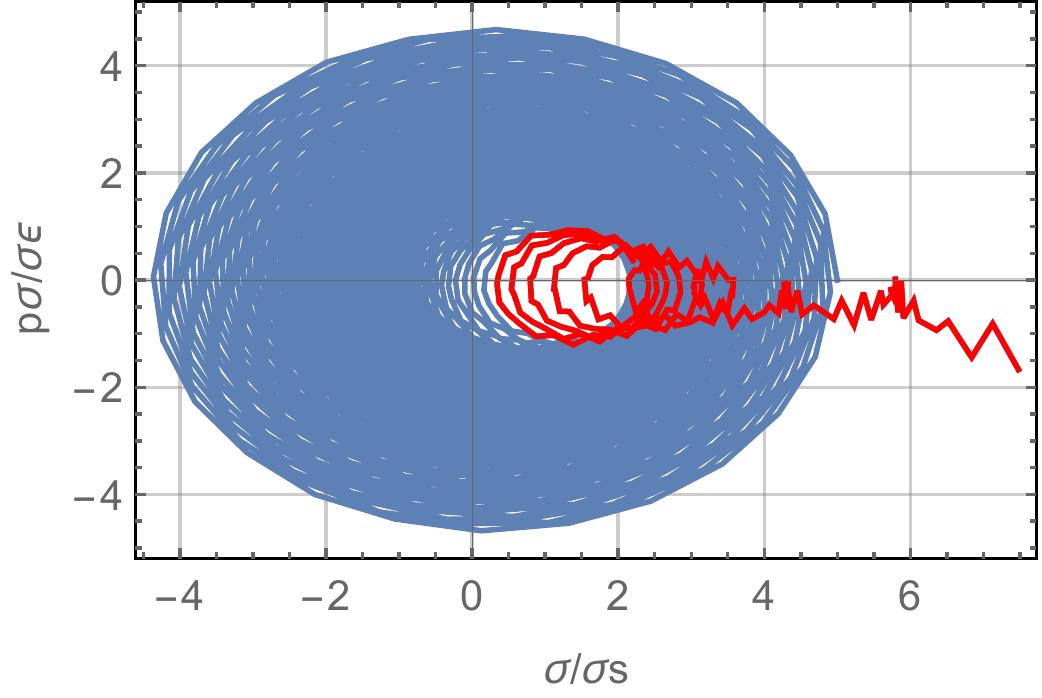}
\caption{Longitudinal phase trajectories from numerical solution of \eqref{eq:MotionEquation-Sigma-2} and \eqref{eq:MotionEquation-PSigma-2} with vertical action in the form \eqref{eq:Solution-Jz-1} corresponding to initial condition $y=58\sigma_y$. The last 200 turns are shown in red. Compare with bottom rigt plot of Figure \ref{fig:UnstablePhaseTrajectoriesY-1}.}
\label{fig:UnstablePhaseTrajectoriesCalcY-1}
\end{figure}

Resemblance of longitudinal phase trajectories on Figure \ref{fig:UnstablePhaseTrajectoriesY-1} and Figure \ref{fig:UnstablePhaseTrajectoriesCalcY-1} proves our approach in solving longitudinal equations \eqref{eq:MotionEquation-Sigma-2} and \eqref{eq:MotionEquation-PSigma-2}.
Figure \ref{fig:UnstablePhaseTrajectoriesCalcY-1} presents numerical solution of the longitudinal equations \eqref{eq:MotionEquation-Sigma-2} and \eqref{eq:MotionEquation-PSigma-2} with vertical action in the form \eqref{eq:Solution-Jz-1} corresponding to initial condition $y=58\sigma_y$.

\begin{figure}[!ptbh]
\centering
\includegraphics*[width=.48\columnwidth,trim=195 6 190 151, clip]{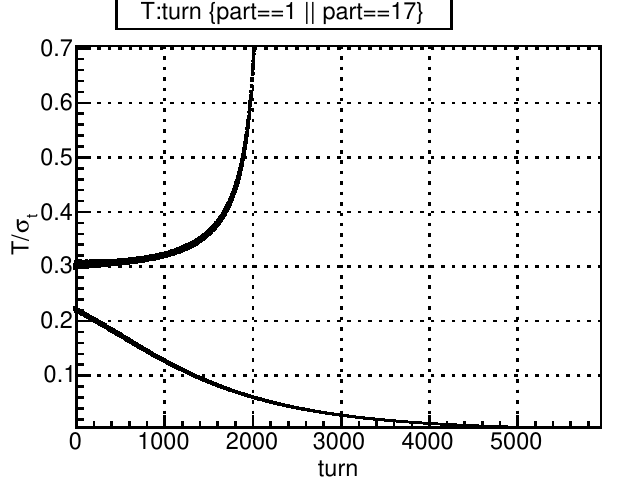}
\hfill
\includegraphics*[width=.48\columnwidth,trim=0 0 0 0, clip]{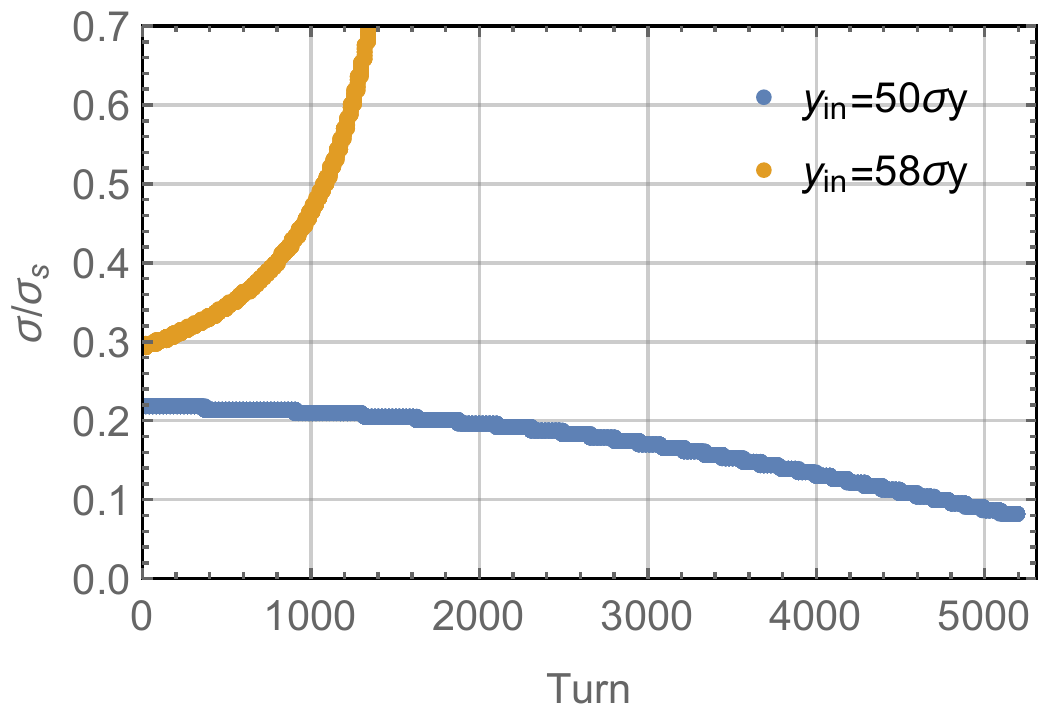}
\\
\caption{Evolution of longitudinal coordinate from tracking (left) and from calculations by \eqref{eq:SynchroSigmaY} and \eqref{eq:SynchroPSigmaY} (right) corresponding to initial conditions $y=50\sigma_y$ and $y=58\sigma_y$ and adjusted longitudinal initial conditions \eqref{eq:SynchroSigmaY} and \eqref{eq:SynchroPSigmaY}.}
\label{fig:SynchronousPhaseTrackY-1}
\end{figure}

Figure \ref{fig:SynchronousPhaseTrackY-1} compares results of tracking and calculations of longitudinal coordinate evolution (synchronous phase) when initial longitudinal conditions were adjusted according to \eqref{eq:SynchroPSigmaY} and \eqref{eq:SynchroSigmaY} in order to eliminate synchrotron oscillations, for two particles with $y=50\sigma_y$ and $y=58\sigma_y$.

\begin{figure}[!ptbh]
\centering
\includegraphics*[width=.48\columnwidth,trim=0 0 0 0, clip]{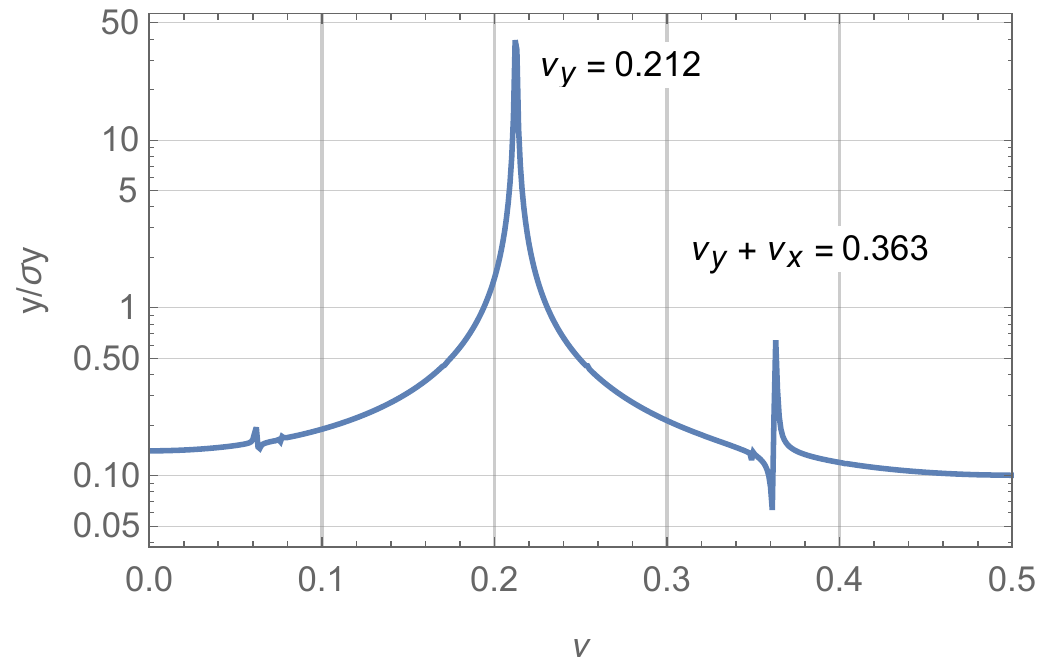}
\hfill
\includegraphics*[width=.48\columnwidth,trim=0 0 0 0, clip]{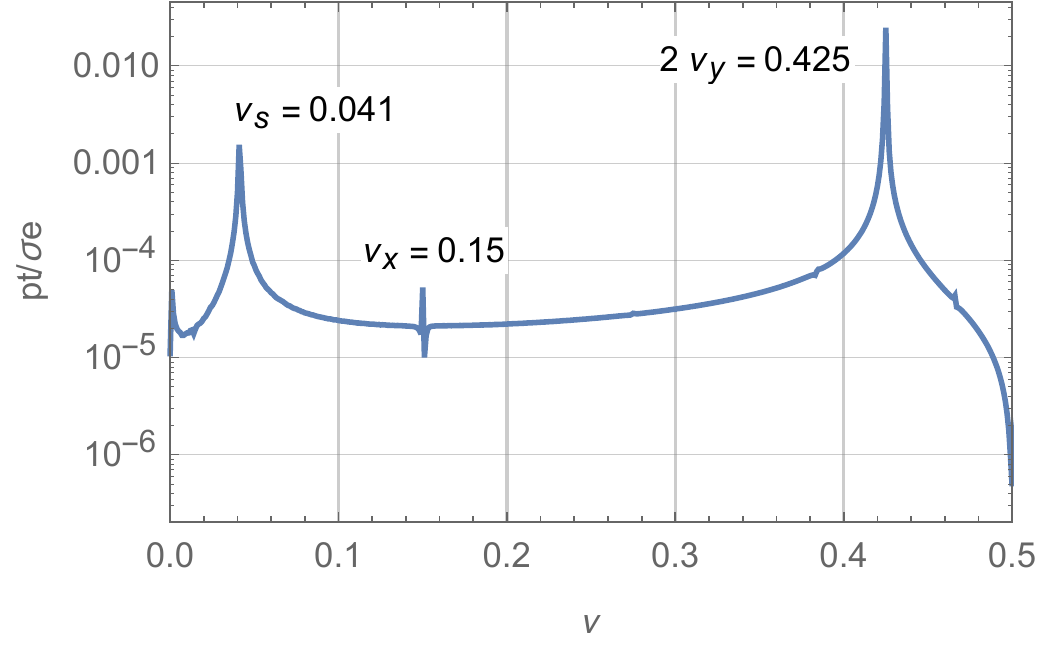}
\\
\caption{Spectrum of vertical (left) and longitudinal (right) motion from tracking corresponding to initial condition $y=58\sigma_y$, and adjusted longitudinal initial conditions \eqref{eq:SynchroSigmaY} and \eqref{eq:SynchroPSigmaY}.}
\label{fig:SpectrumY-1}
\end{figure}

Figure \ref{fig:SpectrumY-1} shows spectra of vertical and longitudinal motion, proving existence of fractional part of double betatron frequency in longitudinal motion. The double frequency harmonic amplitude according to \eqref{eq:Solution-PSigma-4} is $p_\sigma=2.8\times 10^{-2}\sigma_\delta$, which closely corresponds to the value $p_\sigma=2.4\times 10^{-2}\sigma_\delta$ on the right plot of Figure \ref{fig:SpectrumY-1}.

\begin{figure}[!htb]
\centering
\includegraphics*[width=.95\columnwidth,trim=195 154 190 3, clip]{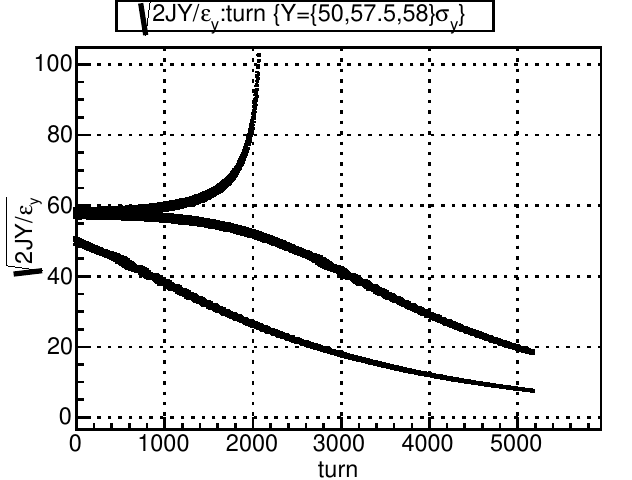}
\caption{Evolution of normalized square root of vertical action from tracking corresponding to initial conditions $y=50\sigma_y$, $y=57.5\sigma_y$, $y=58\sigma_y$, and adjusted longitudinal initial conditions \eqref{eq:SynchroSigmaY} and \eqref{eq:SynchroPSigmaY}.}
\label{fig:JYtrack-1}
\end{figure}

\begin{figure}[!htb]
\centering
\includegraphics*[width=.95\columnwidth,trim=0 0 0 0, clip]{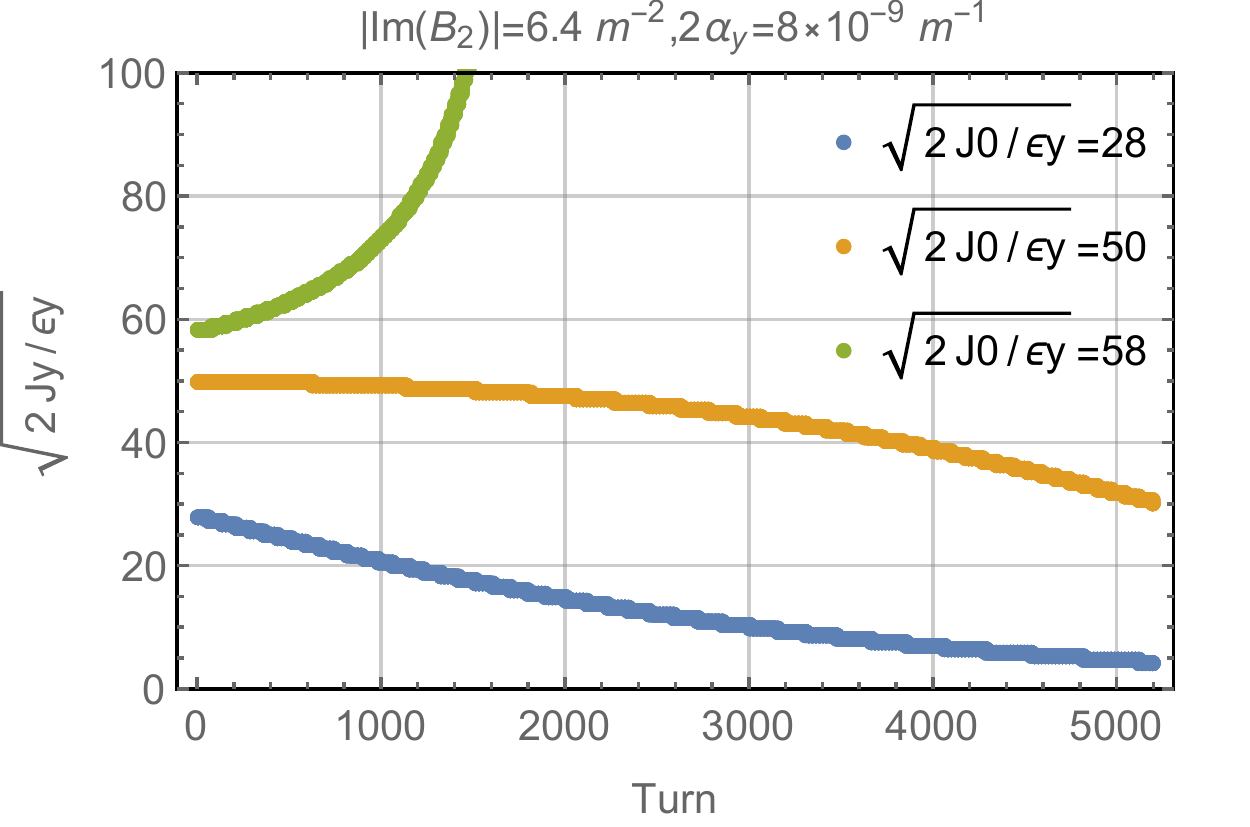}
\caption{Evolution of normalized square root of vertical action from tracking corresponding to initial conditions $y=28\sigma_y$, $y=50\sigma_y$, $y=58\sigma_y$.}
\label{fig:JYcalc-1}
\end{figure}

Figure \ref{fig:JYtrack-1} and Figure \ref{fig:JYcalc-1} compare vertical action evolution from tracking and calculation with 
\eqref{eq:Solution-Jz-1}. The boundary of stable motion is $57.5\sigma_y$ from tracking and $52.6\sigma_y$ from calculations by \eqref{eq:Jzlimit}.

\subsection{Horizontal motion}
\begin{figure}[!ptbh]
\centering
\includegraphics*[width=.48\columnwidth,trim=0 0 0 0, clip]{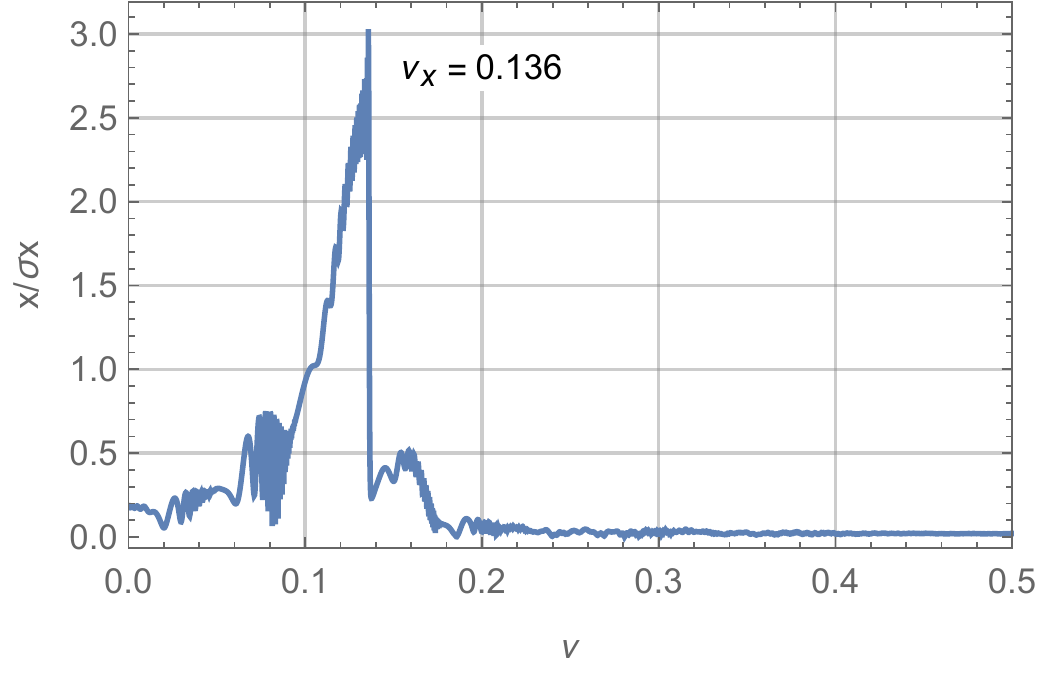}
\hfill
\includegraphics*[width=.48\columnwidth,trim=0 0 0 0, clip]{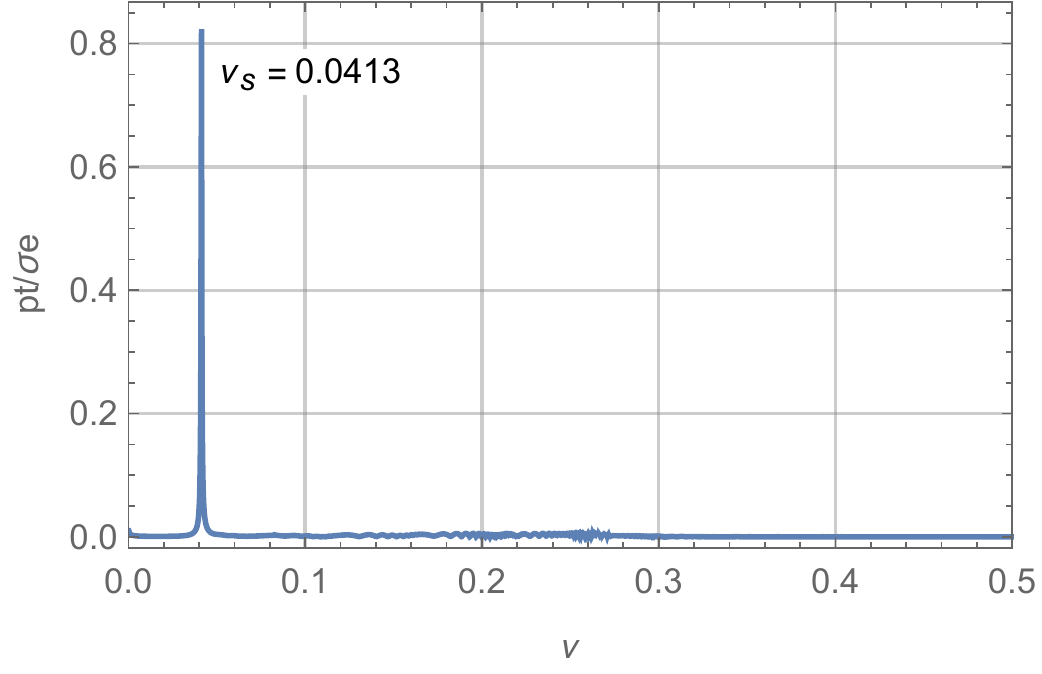}
\\
\caption{Spectrum of horizontal (left) and longitudinal (right) motion from tracking corresponding to initial condition $x=95.5\sigma_x$, and adjusted longitudinal initial conditions. The longitudinal harmonic at double betatron frequency is too small to be observed.}
\label{fig:SpectrumX-1}
\end{figure}

For given horizontal tune harmonic number is $n=-538$, $\mbox{\ae}_x=1.8\times 10^{-5}$~m$^{-1}$, $k_s=2.6\times 10^{-6}$~m$^{-1}$. The harmonics \eqref{eq:harmonicF}, \eqref{eq:harmonicP} and  \eqref{eq:harmonicC} are
\begin{align*}
F_{x,n}&=(-0.003, -1.5\times 10^{-5})\,\mbox{m}^{-3}	& \left|F_{x,n}\right|&=0.003\,\mbox{m}^{-3} \\
P_{x,n}&=(-0.004,-5\times 10^{-4})\,\mbox{m}^{-1}		& \left|P_{x,n}\right|&=0.004\,\mbox{m}^{-1} \\
c_n&=(-2.15,-214)\,\mbox{m}^{-1}									& \left|c_n\right|&=214\,\mbox{m}^{-1}\,.
\end{align*}
The harmonic $\left|c_n\right|$ for horizontal motion is about 30 times smaller than for vertical; therefore, modulation of the longitudinal motion happens at larger amplitudes, which are already unstable due to nonlinear dynamics. This is proven by spectra of horizontal and vertical motion for particle with initial condition $x=95.5\sigma_x$ on Figure \ref{fig:SpectrumX-1}.

\section{Conclusion}
In horizontal plane, additional energy loss due to radiation in quadrupoles, shifts synchronous point and develops large synchrotron oscillations. Horizontal betatron tune dependence on amplitude and chromaticity of this detuning shift the tune toward the integer resonance resulting in particle loss. This is similar to Radiative Beta-Synchrotron Coupling (RBSC) proposed by Jowett \cite{Jowett:1994yt}.

Dynamic aperture reduction in the vertical plane with inclusion of synchrotron radiation in quadrupoles in FCC-ee is due to parametric resonance with modulation amplitude dependent on the square of oscillation amplitude. Radiation from quadrupoles modulates the particle energy at the double betatron frequency; therefore, quadrupole focusing strength also varies at the doubled betatron frequency creating the resonant condition. However, due to friction, resonance develops only if oscillation amplitude is larger than a certain value. The remarkable property of this resonance is that it occurs at any betatron tune (not exactly at half-integer) and, hence, can be labeled as ``self-inducing parametric resonance”. Our calculations give the border of dynamic aperture $R_y=52.6\sigma_y$, which corresponds well to the tracking result $R_y=57\sigma_y$.

\begin{acknowledgments}
We wish to thank John Jowett for his works on electron dynamics with synchrotron radiation, which educated us and helped to accomplish this study.

We are thankful to Katsunobu Oide for the FCC-ee lattice, collaboration and expressed interest to the present work.

We are grateful to Eugene Perevedentsev and Nikolay Vinokurov for reading the manuscript and valuable comments.
\end{acknowledgments}

\appendix

\section{Parametric resonance without damping and amplitude independent modulation}
\begin{figure}[!htb]
\centering
\includegraphics*[width=.9\columnwidth,trim=0 0 0 0, clip]{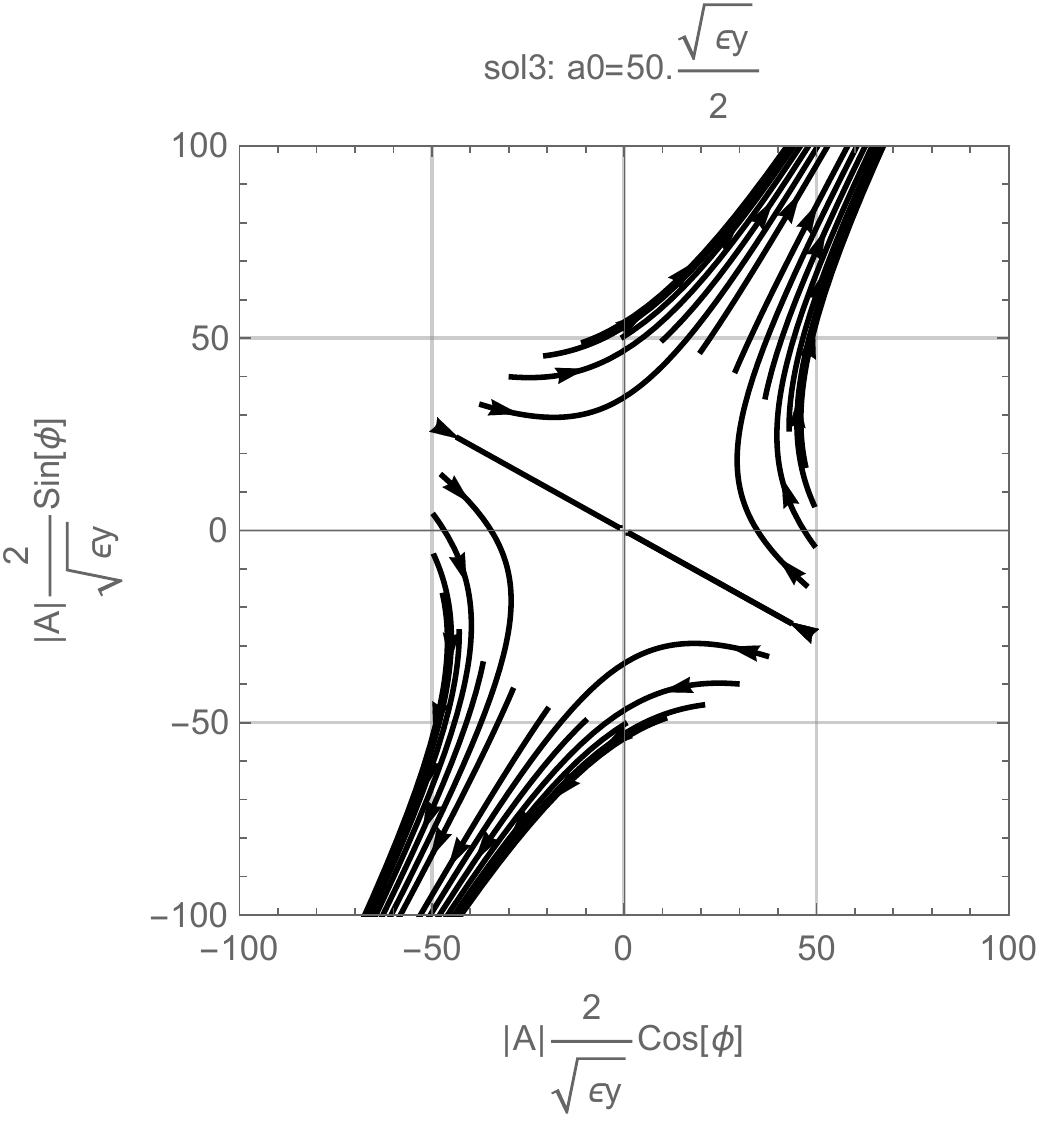}
\caption{Evolution of the average particle trajectories, solution of equations \eqref{eq:Az-6} and \eqref{eq:Phiz-6} with the same initial amplitude and different initial phases. Initial amplitude corresponds to  $y0(\varphi_y=0)=50\sigma_y$}
\label{fig:ParametricDampOff}
\end{figure}

Considering truncated equation \eqref{eq:zdoubleprime-1} without damping
\begin{equation}
\label{eq:zdoubleprime-2}
y''-\left(K_1-(K_1-K_2\eta)p_\sigma\right)y=0\,,
\end{equation}
where modulation does not depends on the amplitude
\begin{equation}
\label{eq:Solution-PSigma-5}
p_\sigma=
g_n e^{i\mbox{\ae}_y s}+g_n^*e^{-i\mbox{\ae}_y s}
=2\left|g_n\right|\cos\left(\mbox{\ae}_y s+\chi_0\right)\,,
\end{equation}
and $\chi_0=\arg(g_n)$, $g_n=c_n (50\sqrt{\varepsilon_y}/2)^2=const$. Now, equation \eqref{eq:zdoubleprime-2} describes a usual parametric resonance with exact resonance condition $\mbox{\ae}_y=\{2\nu_y\}$ . The averaged equations are
\begin{align}
\label{eq:Az-5}
A_y&=i B_2 A_y^*\,, \\
\label{eq:Az-6}
a_y'&=-a_y\left|B_2\right|\sin(-2\varphi_y+\varphi_2)\,, \\
\label{eq:Phiz-6}
\varphi_y'&=\left|B_2\right|\cos(-2\varphi_y+\varphi_2)\,,
\end{align}
where 
\begin{equation}
\label{eq:B2-5}
B_2=\frac{1}{2}g_n\left<(K_1-K_2\eta)\beta_y e^{i(-2\psi_y+\mbox{\ae}_y s)}\right>\,,
\end{equation}
and $A_y=a_y e^{i\varphi_y}$, $B_2=\left|B_2\right|e^{i\varphi_2}$.
Equations \eqref{eq:Az-6} and \eqref{eq:Phiz-6} have two stable points with $\varphi_y'=0$
\begin{equation}
\varphi_y=\frac{\varphi_2}{2}\pm \frac{\pi}{4} +\pi n\,,
\end{equation}
where $n$ is integer. At these points the modulus of the amplitude is
\begin{equation}
a_y(s)=a_{y,0}e^{\pm\left|B_2\right|s}\,.
\end{equation}
The Figure \ref{fig:ParametricDampOff} shows numerical solution of equations \eqref{eq:Az-6} and \eqref{eq:Phiz-6} on the plane of the average particle trajectories $y/\sigma_y=2\left|A_y\right|\cos(\varphi_y)/\sqrt{\varepsilon_y}$ and $p_y/\sigma_{py}=2\left|A_y\right|\sin(\varphi_y)/\sqrt{\varepsilon_y}$, where initial conditions were $a_y(0)=50\sqrt{\varepsilon_y}/2$ and $\varphi_y$ is uniformly distributed between $(0;2\pi)$. As expected,  all trajectories are diverging.

\section{Parametric resonance with damping and amplitude independent modulation}
\begin{figure}[!phtb]
\centering
\includegraphics*[width=.9\columnwidth,trim=0 0 0 0, clip]{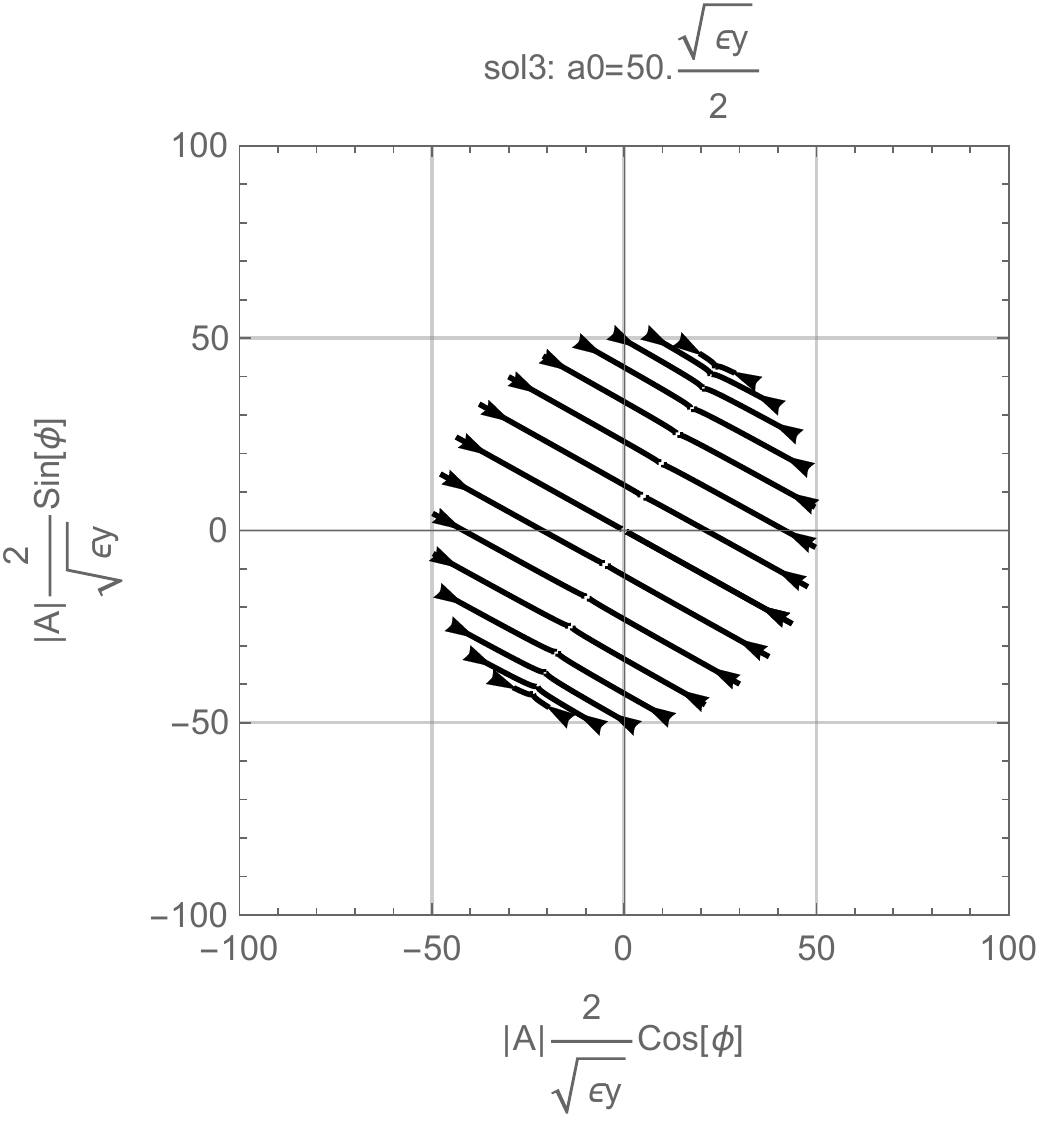}
\caption{Evolution of the average particle trajectories, solution of equations \eqref{eq:Az-6} and \eqref{eq:Phiz-6} with the same initial amplitude and different initial phases. Initial amplitude corresponds to  $y0(\varphi_y=0)=50\sigma_y$, with small modulation amplitude.}
\label{fig:ParametricDampOnSmall}
\end{figure}

\begin{figure}[!phtb]
\centering
\includegraphics*[width=.9\columnwidth,trim=0 0 0 0, clip]{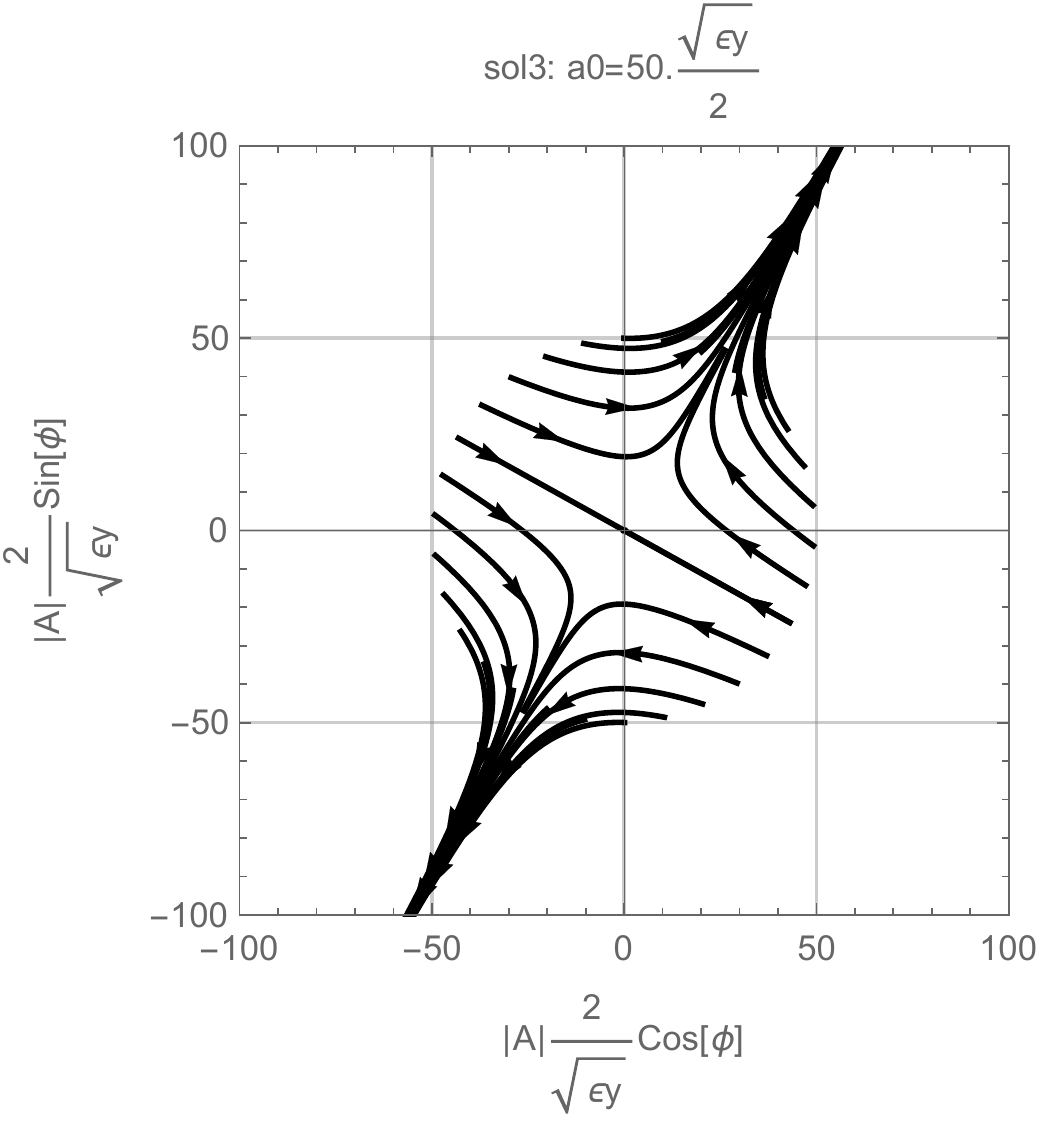}
\caption{Evolution of the average particle trajectories, solution of equations \eqref{eq:Az-6} and \eqref{eq:Phiz-6} with the same initial amplitude and different initial phases. Initial amplitude corresponds to  $y0(\varphi_y=0)=50\sigma_y$, with large modulation amplitude.}
\label{fig:ParametricDampOn}
\end{figure}

Adding the damping term in the equation of the vertical motion yields
\begin{equation}
\label{eq:zdoubleprime-3}
y''-\left(K_1-(K_1-K_2\eta)p_\sigma\right)y+\Gamma K_0^2 y'=0\,.
\end{equation}
The averaged equations are
\begin{align}
\label{eq:Az-7}
A_y&=-B_1 A_y+i B_2 A_y^*\,, \\
\label{eq:Az-8}
a_y'&=-a_y\left|B_1\right|\cos(\varphi_1)-a_y\left|B_2\right|\sin(-2\varphi_y+\varphi_2)\,, \\
\label{eq:Phiz-8}
\varphi_y'&=-\left|B_1\right|\sin(\varphi_1)+\left|B_2\right|\cos(-2\varphi_y+\varphi_2)\,,
\end{align}
where 
\begin{align}
\label{eq:B1-6}
B_1&=\frac{1}{2}\left<\Gamma K_0^2(1+i\alpha_y)\right> \\
\label{eq:B2-6}
B_2&=\frac{1}{2}g_n\left<(K_1-K_2\eta)\beta_y e^{i(-2\psi_y+\mbox{\ae}_y s)}\right>\,,
\end{align}
and $A_y=a_y e^{i\varphi_y}$, $B_1=\left|B_1\right|e^{i\varphi_1}$, $B_2=\left|B_2\right|e^{i\varphi_2}$.
Neglecting $\left|B_1\right|\sin(\varphi_1)$ equations \eqref{eq:Az-8} and \eqref{eq:Phiz-8} have the same two stable points with $\varphi_y'=0$
\begin{equation}
\varphi_y=\frac{\varphi_2}{2}\pm \frac{\pi}{4} +\pi n\,,
\end{equation}
where $n$ is integer. At these points the modulus of the amplitude is
\begin{equation}
a_y(s)=a_{y,0}e^{-\left|B_1\right|\cos(\varphi_1) \pm\left|B_2\right|s}\,.
\end{equation}
The Figures \ref{fig:ParametricDampOnSmall} and \ref{fig:ParametricDampOn} show numerical solution of equations \eqref{eq:Az-8} and \eqref{eq:Phiz-8} on the plane of the average particle trajectories $y/\sigma_y=2\left|A_y\right|\cos(\varphi_y)/\sqrt{\varepsilon_y}$ and $p_y/\sigma_{py}=2\left|A_y\right|\sin(\varphi_y)/\sqrt{\varepsilon_y}$, where initial conditions were $a_y(0)=50\sqrt{\varepsilon_y}/2$ and $\varphi_y$ is uniformly distributed between $(0;2\pi)$. Because of damping we have different behavior depending the strength of the modulation amplitude: if modulation amplitude is small then all trajectories are stable (FIG. \ref{fig:ParametricDampOnSmall}), if modulation amplitude is large then all trajectories are diverging  (FIG. \ref{fig:ParametricDampOn}).

\section{Parametric resonance with damping and amplitude dependent modulation}
\begin{figure}[!ptbh]
\centering
\includegraphics*[width=.9\columnwidth,trim=0 0 0 0, clip]{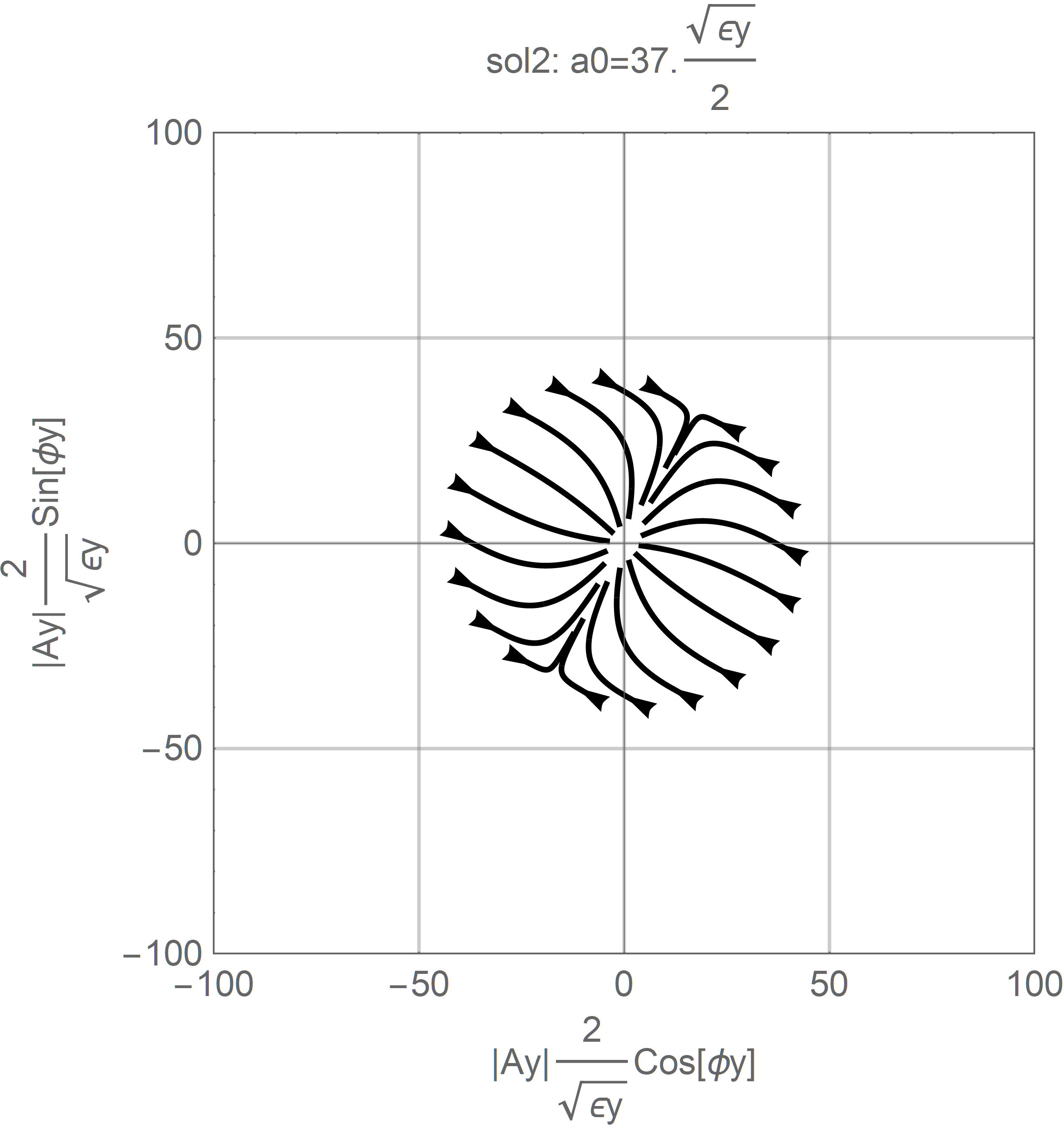}
\caption{Evolution of the average particle trajectories, solution of equations \eqref{eq:Az-4} and \eqref{eq:Phiz-4} with the same initial amplitude and different initial phases. Initial amplitude corresponds to  $y0(\varphi_y=0)=37\sigma_y$}
\label{fig:ParametricRealDampOn-1}
\end{figure}

\begin{figure}[!ptbh]
\centering
\includegraphics*[width=.9\columnwidth,trim=0 0 0 0, clip]{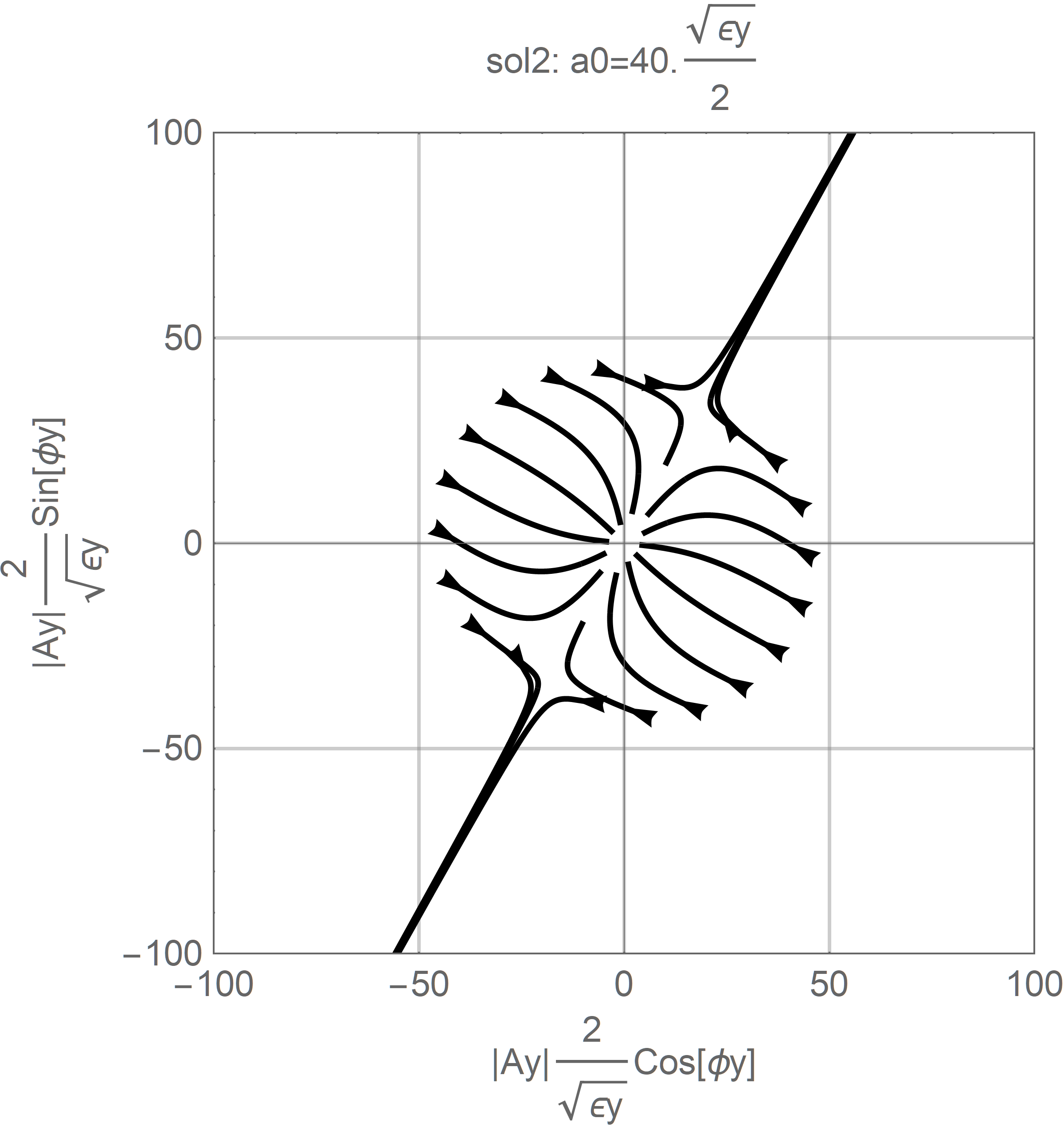}
\caption{Evolution of the average particle trajectories, solution of equations \eqref{eq:Az-4} and \eqref{eq:Phiz-4} with the same initial amplitude and different initial phases. Initial amplitude corresponds to  $y0(\varphi_y=0)=40\sigma_y$}
\label{fig:ParametricRealDampOn-2}
\end{figure}

\begin{figure}[!ptbh]
\centering
\includegraphics*[width=.9\columnwidth,trim=0 0 0 0, clip]{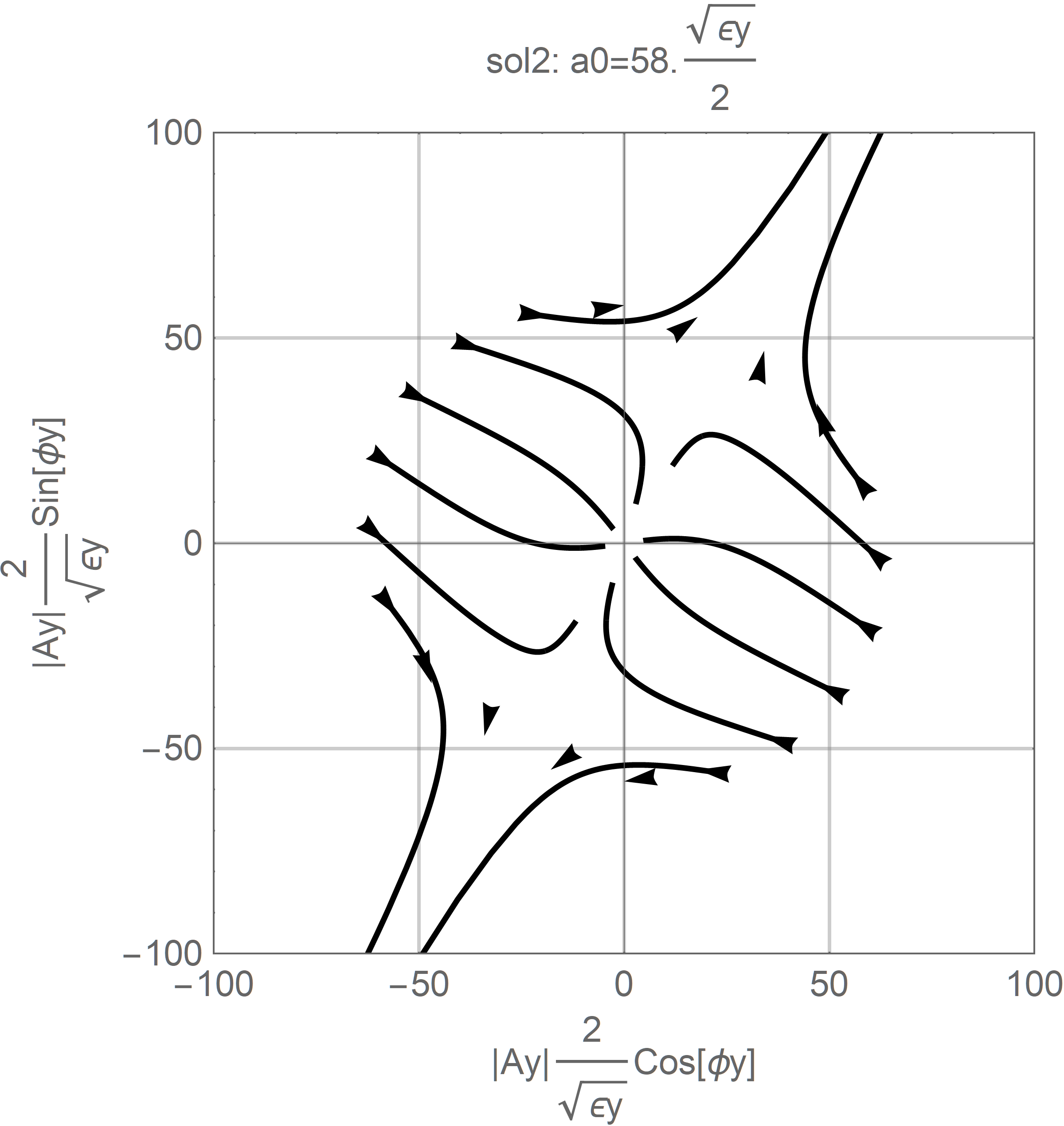}
\caption{Evolution of the average particle trajectories, solution of equations \eqref{eq:Az-4} and \eqref{eq:Phiz-4} with the same initial amplitude and different initial phases. Initial amplitude corresponds to  $y0(\varphi_y=0)=58\sigma_y$}
\label{fig:ParametricRealDampOn-3}
\end{figure}

In the realistic case of equation \eqref{eq:zdoubleprime-1} with coefficients \eqref{eq:B1} and \eqref{eq:B2}, the modulation amplitude depends on the square of the oscillation amplitude. Therefore, depending on initial amplitude either all trajectories are stable, or some are stable and others are unstable, or all unstable. Figures \ref{fig:ParametricRealDampOn-1} (all trajectories are stable), \ref{fig:ParametricRealDampOn-2} (some trajectories are unstable) and \ref{fig:ParametricRealDampOn-3} (majority of trajectories are unstable) show numerical solution of equations \eqref{eq:Az-4} and \eqref{eq:Phiz-4} on the plane of the average particle trajectories $y/\sigma_y=2\left|A_y\right|\cos(\varphi_y)/\sqrt{\varepsilon_y}$ and $p_y/\sigma_{py}=2\left|A_y\right|\sin(\varphi_y)/\sqrt{\varepsilon_y}$, with three different initial amplitudes and uniformly distributed $\varphi_y$ between $(0;2\pi)$. All trajectories are stable for $y0(\varphi_y=0)=37\sigma_y$, and with larger initial amplitude number of unstable trajectories increases. 

\clearpage

\bibliography{References.bib}

\begin{thebibliography}{17}%
\makeatletter
\providecommand \@ifxundefined [1]{%
 \@ifx{#1\undefined}
}%
\providecommand \@ifnum [1]{%
 \ifnum #1\expandafter \@firstoftwo
 \else \expandafter \@secondoftwo
 \fi
}%
\providecommand \@ifx [1]{%
 \ifx #1\expandafter \@firstoftwo
 \else \expandafter \@secondoftwo
 \fi
}%
\providecommand \natexlab [1]{#1}%
\providecommand \enquote  [1]{``#1''}%
\providecommand \bibnamefont  [1]{#1}%
\providecommand \bibfnamefont [1]{#1}%
\providecommand \citenamefont [1]{#1}%
\providecommand \href@noop [0]{\@secondoftwo}%
\providecommand \href [0]{\begingroup \@sanitize@url \@href}%
\providecommand \@href[1]{\@@startlink{#1}\@@href}%
\providecommand \@@href[1]{\endgroup#1\@@endlink}%
\providecommand \@sanitize@url [0]{\catcode `\\12\catcode `\$12\catcode
  `\&12\catcode `\#12\catcode `\^12\catcode `\_12\catcode `\%12\relax}%
\providecommand \@@startlink[1]{}%
\providecommand \@@endlink[0]{}%
\providecommand \url  [0]{\begingroup\@sanitize@url \@url }%
\providecommand \@url [1]{\endgroup\@href {#1}{\urlprefix }}%
\providecommand \urlprefix  [0]{URL }%
\providecommand \Eprint [0]{\href }%
\providecommand \doibase [0]{http://dx.doi.org/}%
\providecommand \selectlanguage [0]{\@gobble}%
\providecommand \bibinfo  [0]{\@secondoftwo}%
\providecommand \bibfield  [0]{\@secondoftwo}%
\providecommand \translation [1]{[#1]}%
\providecommand \BibitemOpen [0]{}%
\providecommand \bibitemStop [0]{}%
\providecommand \bibitemNoStop [0]{.\EOS\space}%
\providecommand \EOS [0]{\spacefactor3000\relax}%
\providecommand \BibitemShut  [1]{\csname bibitem#1\endcsname}%
\let\auto@bib@innerbib\@empty
\bibitem [{\citenamefont {FCC}()}]{FCC1}%
  \BibitemOpen
  \bibfield  {author} {\bibinfo {author} {\bibnamefont {FCC}},\ }\href
  {http://cern.ch/fcc} {\emph {\bibinfo {title} {http://cern.ch/fcc}}},\
  \bibinfo {type} {Tech. Rep.}\BibitemShut {Stop}%
\bibitem [{\citenamefont {CEPC}()}]{CEPC1}%
  \BibitemOpen
  \bibfield  {author} {\bibinfo {author} {\bibnamefont {CEPC}},\ }\href
  {http://cepc.ihep.ac.cn} {\emph {\bibinfo {title}
  {http://cepc.ihep.ac.cn}}},\ \bibinfo {type} {Tech. Rep.}\BibitemShut {Stop}%
\bibitem [{\citenamefont {Augustin}\ \emph {et~al.}(1978)\citenamefont
  {Augustin}, \citenamefont {Dikansky}, \citenamefont {Derbenev}, \citenamefont
  {Rees}, \citenamefont {Richter}, \citenamefont {Skrinsky}, \citenamefont
  {Tigner},\ and\ \citenamefont {Wiedemann}}]{Augustin:1978ah}%
  \BibitemOpen
  \bibfield  {author} {\bibinfo {author} {\bibfnamefont {J.~E.}\ \bibnamefont
  {Augustin}}, \bibinfo {author} {\bibfnamefont {N.}~\bibnamefont {Dikansky}},
  \bibinfo {author} {\bibfnamefont {{\relax Ya}.}~\bibnamefont {Derbenev}},
  \bibinfo {author} {\bibfnamefont {J.}~\bibnamefont {Rees}}, \bibinfo {author}
  {\bibfnamefont {B.}~\bibnamefont {Richter}}, \bibinfo {author} {\bibfnamefont
  {A.}~\bibnamefont {Skrinsky}}, \bibinfo {author} {\bibfnamefont
  {M.}~\bibnamefont {Tigner}}, \ and\ \bibinfo {author} {\bibfnamefont
  {H.}~\bibnamefont {Wiedemann}},\ }\bibfield  {booktitle} {\emph {\bibinfo
  {booktitle} {{PROCEEDINGS OF THE WORKSHOP ON POSSIBILITIES AND LIMITATIONS OF
  ACCELERATORS AND DETECTORS, HELD AT FERMI NATIONAL ACCELERATOR LABORATORY,
  OCTOBER 15-21, 1978}}},\ }\href@noop {} {\bibfield  {journal} {\bibinfo
  {journal} {eConf}\ }\textbf {\bibinfo {volume} {C781015}},\ \bibinfo {pages}
  {009} (\bibinfo {year} {1978})}\BibitemShut {NoStop}%
\bibitem [{\citenamefont {Telnov}(2013)}]{Telnov:2012rm}%
  \BibitemOpen
  \bibfield  {author} {\bibinfo {author} {\bibfnamefont {V.~I.}\ \bibnamefont
  {Telnov}},\ }\href {\doibase 10.1103/PhysRevLett.110.114801} {\bibfield
  {journal} {\bibinfo  {journal} {Phys. Rev. Lett.}\ }\textbf {\bibinfo
  {volume} {110}},\ \bibinfo {pages} {114801} (\bibinfo {year} {2013})},\
  \Eprint {http://arxiv.org/abs/1203.6563} {arXiv:1203.6563 [physics.acc-ph]}
  \BibitemShut {NoStop}%
\bibitem [{\citenamefont {Bogomyagkov}\ \emph {et~al.}(2014)\citenamefont
  {Bogomyagkov}, \citenamefont {Levichev},\ and\ \citenamefont
  {Shatilov}}]{Bogomyagkov:2013uja}%
  \BibitemOpen
  \bibfield  {author} {\bibinfo {author} {\bibfnamefont {A.}~\bibnamefont
  {Bogomyagkov}}, \bibinfo {author} {\bibfnamefont {E.}~\bibnamefont
  {Levichev}}, \ and\ \bibinfo {author} {\bibfnamefont {D.}~\bibnamefont
  {Shatilov}},\ }\href {\doibase 10.1103/PhysRevSTAB.17.041004} {\bibfield
  {journal} {\bibinfo  {journal} {Phys. Rev. ST Accel. Beams}\ }\textbf
  {\bibinfo {volume} {17}},\ \bibinfo {pages} {041004} (\bibinfo {year}
  {2014})},\ \Eprint {http://arxiv.org/abs/1311.1580} {arXiv:1311.1580
  [physics.acc-ph]} \BibitemShut {NoStop}%
\bibitem [{\citenamefont {Jowett}(1994)}]{Jowett:1994yt}%
  \BibitemOpen
  \bibfield  {author} {\bibinfo {author} {\bibfnamefont {J.}~\bibnamefont
  {Jowett}},\ }\bibfield  {booktitle} {\emph {\bibinfo {booktitle}
  {{Proceedings: LEP Performance Workshop, 4th, Chamonix, France, Jan, 17-21,
  1994}}},\ }\href@noop {} {\bibfield  {journal} {\bibinfo  {journal} {Conf.
  Proc.}\ }\textbf {\bibinfo {volume} {C9401174}},\ \bibinfo {pages} {47}
  (\bibinfo {year} {1994})},\ \bibinfo {note} {[,47(1994)]}\BibitemShut
  {NoStop}%
\bibitem [{\citenamefont {Barbarin}\ \emph {et~al.}(1994)\citenamefont
  {Barbarin}, \citenamefont {Iselin},\ and\ \citenamefont
  {Jowett}}]{Barbarin:1994gy}%
  \BibitemOpen
  \bibfield  {author} {\bibinfo {author} {\bibfnamefont {F.}~\bibnamefont
  {Barbarin}}, \bibinfo {author} {\bibfnamefont {F.~C.}\ \bibnamefont
  {Iselin}}, \ and\ \bibinfo {author} {\bibfnamefont {J.~M.}\ \bibnamefont
  {Jowett}},\ }\bibfield  {booktitle} {\emph {\bibinfo {booktitle} {{4th
  European Particle Accelerator Conference (EPAC 94) London, England, June
  27-July 1, 1994}}},\ }\href@noop {} {\bibfield  {journal} {\bibinfo
  {journal} {Conf. Proc.}\ }\textbf {\bibinfo {volume} {C940627}},\ \bibinfo
  {pages} {193} (\bibinfo {year} {1994})}\BibitemShut {NoStop}%
\bibitem [{\citenamefont {Jowett}(1998)}]{Jowett:1998au}%
  \BibitemOpen
  \bibfield  {author} {\bibinfo {author} {\bibfnamefont {J.~M.}\ \bibnamefont
  {Jowett}},\ }in\ \href@noop {} {\emph {\bibinfo {booktitle} {{Beam dynamics
  issues for e+ e- factories. Proceedings, Advanced ICFA Workshop, ICFA'97,
  Frascati, Italy, October 20-25, 1997}}}}\ (\bibinfo {year} {1998})\ pp.\
  \bibinfo {pages} {15--38}\BibitemShut {NoStop}%
\bibitem [{\citenamefont {SAD}()}]{SAD}%
  \BibitemOpen
  \bibfield  {author} {\bibinfo {author} {\bibnamefont {SAD}},\ }\href
  {http://acc-physics.kek.jp/SAD/index.html} {\emph {\bibinfo {title}
  {http://acc-physics.kek.jp/SAD/index.html}}},\ \bibinfo {type} {Tech.
  Rep.}\BibitemShut {Stop}%
\bibitem [{\citenamefont {Oide}\ \emph {et~al.}(2016)\citenamefont {Oide} \emph
  {et~al.}}]{Oide:FCC2016}%
  \BibitemOpen
  \bibfield  {author} {\bibinfo {author} {\bibfnamefont {K.}~\bibnamefont
  {Oide}} \emph {et~al.},\ }\href {\doibase
  10.1103/PhysRevAccelBeams.19.111005, 10.1103/PhysRevAccelBeams.20.049901}
  {\bibfield  {journal} {\bibinfo  {journal} {Phys. Rev. Accel. Beams}\
  }\textbf {\bibinfo {volume} {19}},\ \bibinfo {pages} {111005} (\bibinfo
  {year} {2016})},\ \bibinfo {note} {[Addendum: Phys. Rev. Accel.
  Beams20,no.4,049901(2017)]},\ \Eprint {http://arxiv.org/abs/1610.07170}
  {1610.07170 [physics.acc-ph]} \BibitemShut {NoStop}%
\bibitem [{\citenamefont {MADX}()}]{MADX}%
  \BibitemOpen
  \bibfield  {author} {\bibinfo {author} {\bibnamefont {MADX}},\ }\href
  {http://madx.web.cern.ch/madx} {\emph {\bibinfo {title}
  {http://madx.web.cern.ch/madx}}},\ \bibinfo {type} {Tech. Rep.}\BibitemShut
  {Stop}%
\bibitem [{\citenamefont {Glukhov}\ \emph {et~al.}(2015)\citenamefont {Glukhov}
  \emph {et~al.}}]{TracKing}%
  \BibitemOpen
  \bibfield  {author} {\bibinfo {author} {\bibfnamefont {S.}~\bibnamefont
  {Glukhov}} \emph {et~al.},\ }in\ \href@noop {} {\emph {\bibinfo {booktitle}
  {{Proceedings of ICAP2015, Shanghai, China, 2015}}}}\ (\bibinfo {year}
  {2015})\ pp.\ \bibinfo {pages} {115--117}\BibitemShut {NoStop}%
\bibitem [{\citenamefont {Forest}\ and\ \citenamefont
  {Milutinovic}(1988)}]{Forest:1987dr}%
  \BibitemOpen
  \bibfield  {author} {\bibinfo {author} {\bibfnamefont {E.}~\bibnamefont
  {Forest}}\ and\ \bibinfo {author} {\bibfnamefont {J.}~\bibnamefont
  {Milutinovic}},\ }\href {\doibase 10.1016/0168-9002(88)90123-4} {\bibfield
  {journal} {\bibinfo  {journal} {Nucl. Instrum. Meth.}\ }\textbf {\bibinfo
  {volume} {A269}},\ \bibinfo {pages} {474} (\bibinfo {year}
  {1988})}\BibitemShut {NoStop}%
\bibitem [{\citenamefont {Helm}\ \emph {et~al.}(1973)\citenamefont {Helm},
  \citenamefont {Lee}, \citenamefont {Morton},\ and\ \citenamefont
  {Sands}}]{Helm:1973radintegrals}%
  \BibitemOpen
  \bibfield  {author} {\bibinfo {author} {\bibfnamefont {R.~H.}\ \bibnamefont
  {Helm}}, \bibinfo {author} {\bibfnamefont {M.~J.}\ \bibnamefont {Lee}},
  \bibinfo {author} {\bibfnamefont {P.~L.}\ \bibnamefont {Morton}}, \ and\
  \bibinfo {author} {\bibfnamefont {M.}~\bibnamefont {Sands}},\ }\bibfield
  {booktitle} {\emph {\bibinfo {booktitle} {{Proceedings, 1973 Particle
  Accelerator Conference, Accelerator Engineering and Technology: San
  Francisco, California, March 5-7, 1973}}},\ }\href {\doibase
  10.1109/TNS.1973.4327284} {\bibfield  {journal} {\bibinfo  {journal} {IEEE
  Trans. Nucl. Sci.}\ }\textbf {\bibinfo {volume} {20}},\ \bibinfo {pages}
  {900} (\bibinfo {year} {1973})}\BibitemShut {NoStop}%
\bibitem [{\citenamefont {Courant}\ and\ \citenamefont
  {Snyder}(1958)}]{ref:Courant:1997}%
  \BibitemOpen
  \bibfield  {author} {\bibinfo {author} {\bibfnamefont {E.~D.}\ \bibnamefont
  {Courant}}\ and\ \bibinfo {author} {\bibfnamefont {H.~S.}\ \bibnamefont
  {Snyder}},\ }\href {\doibase 10.1016/0003-4916(58)90012-5} {\bibfield
  {journal} {\bibinfo  {journal} {Annals Phys.}\ }\textbf {\bibinfo {volume}
  {3}},\ \bibinfo {pages} {1} (\bibinfo {year} {1958})},\ \bibinfo {note}
  {[Annals Phys.281,360(2000)]}\BibitemShut {NoStop}%
\bibitem [{\citenamefont {Jowett}(1987)}]{Jowett:1986pc}%
  \BibitemOpen
  \bibfield  {author} {\bibinfo {author} {\bibfnamefont {J.~M.}\ \bibnamefont
  {Jowett}},\ }\bibfield  {booktitle} {\emph {\bibinfo {booktitle}
  {{Proceedings, 1985 SLAC Summer School on the Physics of High-energy Particle
  Accelerators and 1984 U.S. Summer School on High-Energy Particle
  Accelerators: Stanford, California, July 15-26, 1985}}},\ }\href {\doibase
  10.1063/1.36374} {\bibfield  {journal} {\bibinfo  {journal} {AIP Conf.
  Proc.}\ }\textbf {\bibinfo {volume} {153}},\ \bibinfo {pages} {864} (\bibinfo
  {year} {1987})}\BibitemShut {NoStop}%
\bibitem [{\citenamefont {Jowett}(1986)}]{Jowett:1986yx}%
  \BibitemOpen
  \bibfield  {author} {\bibinfo {author} {\bibfnamefont {J.~M.}\ \bibnamefont
  {Jowett}},\ }in\ \href
  {http://www-public.slac.stanford.edu/sciDoc/docMeta.aspx?slacPubNumber=slac-ap-053.html}
  {\emph {\bibinfo {booktitle} {{CERN Accel.School 1985:0570}}}}\ (\bibinfo
  {year} {1986})\BibitemShut {NoStop}%
\end{thebibliography}%

\end{document}